\documentclass[reprint,
superscriptaddress, amsmath,amssymb,
 prl,
]{revtex4-1}
\usepackage{graphicx}
\usepackage{graphics}
\usepackage{footmisc}
\usepackage{latexsym}
\usepackage{amsmath, amsthm, amssymb, wasysym}
\usepackage{xcolor}
\usepackage{epsfig}
\usepackage{mathrsfs}

\DeclareMathOperator{\okr}{{\stackrel{{\scriptscriptstyle{\mathsf{def}}}}{=}}}
\DeclareMathOperator{\D}{d\!}
\DeclareMathOperator{\E}{e} 
\DeclareMathOperator{\I}{i}

\def\ulamek#1#2{\mbox{\normalfont$\frac{#1}{#2}$}}

\begin{document}
\title{Non-Debye relaxations: The ups and downs of the stretched exponential vs Mittag-Leffler's matchings.}

\author{K. G\'orska}
\email{katarzyna.gorska@ifj.edu.pl} 
\author{A. Horzela}
\email{andrzej.horzela@ifj.edu.pl} 
\affiliation{Institute of Nuclear Physics, Polish Academy of Sciences, PL-31342 Krak\'ow, Poland}

\author{K. A. Penson}
\email{karol.penson@sorbonne-universite.fr} 
\affiliation{Laboratorie de Physique Theorique de la Mati\`{e}re Condens\'{e}e (LPTMC),
CNRS UMR 7600, Sorbonne Universit\'{e}, Campus Pierre et Marie Curie, F-75005 Paris, France.}

% Abstract (Do not insert blank lines, i.e. \\) 
\begin{abstract}
Experimental data collected to provide us with information on the course of dielectric relaxation phenomena are got according to two distinct schemes: one can measure either the time decay of depolarization current or use methods of the broadband dielectric spectroscopy. Both sets of data are usually fitted by time or frequency dependent elementary functions which in turn may be analytically transformed among themselves using the Laplace transform and compared each other. This leads to the question on comparability of results got using just mentioned experimental procedures. If we would like to do that in the time domain we have to go beyond widely accepted Kohlrausch-Williams-Watts approximation and get acquainted with description using the Mittag-Leffler functions. To convince the reader that the latter is not difficult to understand we propose to look at the problem from the point of view of objects sitting in the heart of stochastic processes approach to relaxation. These are the characteristic exponents which are read out from the standard non-Debye frequency dependent patterns. Characteristic functions appear to be expressed in terms of elementary functions which asymptotic analysis is simple. This opens new possibility to compare behavior of functions used to describe non-Debye relaxations. Results of such done comparison are fully confirmed by calculations which use the powerful apparatus of the Mittag-Leffler functions.
\\
\\
This work belongs to the special issue ``Fractional Dynamics: Theory and Applications''. 
% Keywords
%\keyword{non-Debye relaxations; characteristic exponents; subordination; Efross theorem}  % List three to ten pertinent keywords specific to the article, 
\end{abstract}

\maketitle

\section{Introduction}

Description of physical phenomena which kinetics is influenced by complexity, disorder or randomness often requires a radical departure from theoretical methods established for analogous, but simpler, phenomena discussed in textbooks of general physics. Such a situation is met when we get interested in study of dielectric relaxations and encounter their time behavior different from the commonly expected exponential decay. Depending on the experimental setup empirical investigation of the relaxation phenomena and collecting the data is  done by measuring either their time behaviour or frequency characteristics, i.e. the experiment provides us with data in the time or in the frequency domains. Typical example of dielectric relaxation phenomena is provided by a dipolar system which approaches the equilibrium being earlier driven out of it, i.e., polarized, by a step or alternating external electric field. Depolarization is usually described in terms of the relaxation or spectral functions \cite{CJFBoettcher96}. The first just mentioned quantity, namely the time dependent relaxation function $n(t)$, counts dipoles surviving depolarization during the time $(0, t)\subset (0, \infty)$ and evolves form $n(0+) = 1$ to $n(\infty) = 0$. The frequency dependent spectral function $\hat{\phi}(\I\!\omega)$, describing diffractive and absorptive effects, results from the analysis of phenomenological data obtained as a response of the system when it is probed by the harmonic electric field. Defined as the normalized ratio of dielectric permittivities $[\hat{\varepsilon}(\I\!\omega) - \varepsilon_{\infty}]/[\varepsilon_{0} - \varepsilon_{\infty}]$, where $\varepsilon_{\infty} = \lim_{\omega\to\infty} \hat{\varepsilon}(\I\!\omega)$ and $\varepsilon_{0} = \lim_{\omega\to 0} \hat{\varepsilon}(\I\!\omega)$, it is complex valued function which analytical properties stem from those obeyed by $\hat{\varepsilon}(\I\!\omega)$. Following standard rules the data obtained in the  time $t$ or in the frequency $\omega$ domains are interrelated by the Laplace transform $\mathscr{L}[-; \cdot]$
\begin{equation}\label{15/07-1}
\widehat{\phi}(\I\!\omega) = 1 - \I\!\omega \widehat{n}(\I\!\omega), \qquad \widehat{n}(\I\!\omega) = \mathscr{L}[n(t); \I\!\omega].
\end{equation}

In typical non-Debye relaxation experiments data measured in the time domain are usually fitted using the stretched exponential or, in physicists' community language, the Kohlrausch-Williams-Watts (KWW) function
\begin{equation}\label{25/07-1}
n_{KWW}(t) = \exp[-(t/\tau)^{\alpha}] 
\end{equation}
with $\alpha > 0$. In what follows we will consider only the case $\alpha\in(0, 1)$ which preserves the interpretation of the stretched exponential as the continuous sum of Debye exponential decays weighted by a probability distribution which appears to belong to the class of  L\'{e}vy stable distributions \cite{Johnston06,HPollard44}. Phenomenological functions usually used to fit the data in the frequency domain, called the standard non-Debye relaxation patterns, are the Cole-Cole (CC), Havriliak-Negami (HN), and Jurlewicz-Weron-Stanislawsky (JWS) models
\begin{multline}\label{25/07-2}
\widehat{\phi}_{CC}(\I\!\omega) = [1 + (\I\!\omega\tau)^{\alpha}]^{-1}, \quad \widehat{\phi}_{HN}(\I\!\omega) = [1 + (\I\!\omega\tau)^{\alpha}]^{-\beta}, \\ \text{and} \quad 
\widehat{\phi}_{JWS}(\I\!\omega) = 1 - [1 + (\I\!\omega\tau)^{-\alpha}]^{-\beta}, 
\end{multline}
where $\alpha, \beta \in (0, 1]$. 
%\footnote{The range of $\alpha$ and $\beta$ ensures that the spectral functions initially found as a fit for $\omega > 0$ can be analytically continued to the upper half plane $\Im\omega > 0$ and satisfy assumptions of \cite[Theorem 2.7]{GGripenberg90}}. 
Notice that the CC spectral function generalizes the Debye case $\widehat{\phi}_{D}(\I\!\omega) = [1 + (\I\!\omega\tau)]^{-1}$ and simultaneously can be obtained from $\widehat{\phi}_{HN}(\I\!\omega)$ or $\widehat{\phi}_{JWS}(\I\!\omega)$ for $\beta = 1$.  Also, if we set $\alpha = 1$ and $\beta\in(0, 1)$ in the HN model then we get the Cole-Davidson (CD) pattern. Important property of relaxation patterns \eqref{25/07-2} is that if transformed to the time domain they all lead to the relaxation functions $n(t)$ expressed in terms of functions belonging to the family of the Mittag-Leffler functions (see Appendix \ref{a0}). Thus we arrive at a two-fold way how to analyse the non-Debye relaxation phenomena - we can take into account their modeling either in terms of the KWW function or to choose the Mittag-Leffler functions.  None of these approaches is preferred by fundametal theoretical arguments and thus it is understood to treat them as challengers whose usefulness is to be determined by comparison with experimental data. In our opinion results of such comparison lead far beyond its  instructive meaning as they may be used to clarify ambiguities coming from difficulties what experiments are facing in asymptotic (short/long time and high/low frequencies) regimes. Thus looking for arguments shedding light on choosing one of the just mentioned different approaches on experimentally observed data is worth attention and more systematic research.    

{We shall begin comparison of the Mittag-Leffler family and KWW matchings with recalling relations  between the KWW and the standard Mittag-Leffler function is responsible for the time behavior of the CC model. While the KWW function has been used in modeling physical processes mainly in the context of relaxations (e.g. the Curie-von Schweidler law) the CC pattern is by no means restricted to this class of phenomena. Taken for real argument, $t\in\mathbb{R}_{+}$, and $\alpha > 0$,  the CC pattern becomes an example of the generalized Cauchy-Lorentz (GCL) distributions used in numerous fields of basic and applied sciences. To attract attention on the utility of the GCL distributions recall that for $\alpha=2$ it found applications in optics long time ago \cite{JCMaxwell52, RKLuneburg66} and much more recently in quantum mechanics \cite{AJMakowski09} where it describes the so-called Maxwell's fish eye problem.  Interesting application, coming from the interface of the basic and applied science, is using the CC pattern in electrochemistry \cite{RTTGettens09, MHaeri11}, bioelectrochemistry \cite{EHernandez17, EHernandez20, EHernandez21a, EHernandez21b} and photovoltaics \cite{EHernandez21c}. Effects of distributed, i.e. non-Debye, relaxation processes, inhomogeneities of the system and possible deviations from the Gaussian diffusion spreading lead to non-ideal interfacial behaviour and cause that to model electrochemical response one has to go beyond simple models of electric circuits including capacity, like e.g. the Randles circuit \footnote{Randles circuit is an equivalent circuit composed of a resistor in series with combination of a resistor and a capacitor in parallel.}, and to necessity of modifying  current-voltage relations introducing into them (sometimes \textit{ad hoc}) additional time dependent factors given by the KWW or GCL functions \cite{MHaeri11}. Recent progress in investigations of electrochemical processes taking place in biological systems has shown that some results coming from the fractional calculus may be useful to push forward understanding of non-ideal interfacial capacitance. Working example is that if the so-called constant phase element  (CPE) is mounted to replace the standard capacitor in the effective circuit then the differential relation which describes capacitor discharging becomes fractional. Thus we leave the realm of exponential decays (also generalized, like the KWW pattern is) and it becomes quite natural that functions characteristic for fractional calculus, like the Mittag-Leffler ones, come into play and replace exponential-like decay laws. Namely such an approach has been presented in investigations \cite{EHernandez17, EHernandez20, EHernandez21a, EHernandez21b} where it has been also noticed that asymptotic properties of the (standard) Mittag-Leffler function for short times agree with those of the KWW function and that the Mittag-Leffler function interpolates between the KWW for short times and the power-like behavior for long times. Among consequences of this property it has been found that the (standard)  Mittag-Leffler function appears useful not only in studies of short time effects characteristic for biochemical processes   \cite{EHernandez20, EHernandez21b} but also in analysis of the long time phenomena occurring in the perovskite solar cells \cite{EHernandez21a, EHernandez21c}. Here we want to turn the readers' attention and emphasize that attempts to fit the data by the (standard) Mittag-Leffler function instead of the KWW decay suggest to try the use of other function belonging to the Mittag-Leffler family, especially if one would be interested in search of matchings working beyond the leading order of small $t$ asymptotics.}   

{Let us suppose that we perform an experiment in which we are able to observe the relaxation process taking place in two samples, each prepared exactly in the same way, having exactly the same structure and put into the same experimental conditions. Measurements performed for the first sample are taken in the time domain and provide us with direct information on the time decay of polarization while for the second sample we collect spectroscopy data which are next transformed to the time domain. In such twin-like experimental setup it arises the question about agreement between the KWW function commonly used to fit the time data and the relaxation function(s) obtained using the Laplace transform of spectral functions Eq. \eqref{25/07-2} where the choice of suitable pattern emerges from the data analysis. Such comparison was first investigated numerically in Refs. \cite{FAlvarez91, FAlvarez93}  for the KWW and HN models as the authors of analysis were unaware of the Laplace transform of the KWW function. Further study of the problem was announced a few years later in \cite{HavriliakHavriliak96}. Currently, having in hands new mathematical tools,  at that time unknown to the vast majority of physicists, we are going to show how to extend these results using contemporary knowledge, coming from sources far beyond the phenomenology, of the relaxation phenomena. Information expected to help us emerges from dynamics and evolution equations which govern the relaxation processes {\cite{RMetzler02, RGarrappa16, KWeron96, AStanislavsky19}}. However, to get such equations from \textit{ab initio} microscopic rules without implementing far going simplifications is extremely difficult, if possible at all. Thus some "effective" theoretical approach has to be used  - in the case of relaxation processes  suitable mathematical tools are provided by approaches rooted in the stochastic processes theory with the crucial role played by methods grown from the concepts of infinitely divisible distributions and subordination. To give very brief explanation - the most important property of nonnegative nondecreasing stochastic processes governed by infinitely divisible probability distributions is that they are uniquely characterized by functions called the characteristic (either Laplace or L\'evy) exponents which carry on all information concerning distributions under consideration. This formalism adopted for studies of the relaxation phenomena leads to an unexpected result which merges basic, mathematical in fact, theory and pure phenomenology - characteristic exponents may be uniquely reconstructed from the knowledge of spectral function, i.e. experimentally obtained relaxation patterns. This provides us with a new tool to compare various schemes describing relaxation processes - as we mentioned a few lines earlier our goal is to study similarities and/or dissimilarities  of the relaxation descriptions based on the KWW and Mittag-Leffler functions. }

{
The content of our paper goes as follows. Sec.\ref{sec1a} involves preliminaries concerning the characteristic exponents and stochastic approach to relaxations. The spectral and characteristic functions of the KWW model are computed in Sec.\ref{sec2}. Knowledge of the characteristic functions relevant for the standard non-Debye relaxation patterns and their asymptotics enables us to judge the challenge which of them is the best candidate to approximate the KWW model - we remark that results presented in Sec. \ref{sec4} are conclusive only for short times. The last section, Sec.\ref{sec5}, summarizes properties of functions belonging to the Mittag-Leffler family, in particular their asymptotics and (fractional) equations which they obey. We collect in one place results of long and often cumbersome calculations in hope that experimentalists will find them useful in analyses of relaxation experiments. The paper is completed by Conclusions section and five appendices directly devoted to mathematical tools used throughout it.}

\section{Characteristic exponents and stochastic description - a brief tutorial}\label{sec1a}

Characteristic exponents, $\widehat{\Psi}(s)$'s, $s>0$, appear as basic objects reflecting properties of nonnegative infinitely divisible stochastic processes $U(\xi)$ parametrized by a nonnegative nondecreasing random variable $\xi$. They are defined by the relation
\begin{equation}\label{27/07-1}
\left\langle\exp{(-sU(\xi))}\right\rangle=\exp{(-\xi\widehat{\Psi}(s))}
\end{equation}
and given by the L\'evy--Khintchine formula \cite[Eq. (1.3)]{Schilling}. Among properties of the characteristic exponents the most essential is that they belong to the class of Bernstein functions (BFs) closely related to the class of completely monotone functions (CMFs), see Appendix \ref{a4}. To make the notions of BFs and CMFs more intuitive one may understand BFs as ``maximally regularly'' increasing positive functions  while CMFs as ``maximally regularly'' decreasing, but still non-negative, ones. Within the subordination approach to relaxation processes the characteristic exponents $\widehat{\Psi}(s)$ are used to construct distribution functions which subordinate the Debye law assumed to depend on irregularly flowing stochastic operational time $\xi$. Subordination is realized by convoluting such  Debye law with some infinitely divisible probability density function (PDF) $g(t,\xi)$ which provides us with the probability density of finding the system at $\xi$ if it is at the instant of time $t$ measured by a laboratory clock. Having this in mind we can write down $n(t)$ as the  integral decomposition \cite{HCFogedby94,ABaule03,AStanislavsky19}      
\begin{multline}\label{26/07-3}
  n(t) = \int_{0}^{\infty}e^{-B(\tau)\, \xi} g(t, \xi)\D\xi, \\ g(t, \xi) = \mathscr{L}^{-1}\Big[\frac{\widehat{\Psi}(s)}{s} \E^{-\xi \widehat{\Psi}(s)}; t\Big]
\end{multline}
where $B(\tau)$ ($B$ in short-hand-notation) denotes the material, time independent, transition rate characterizing the system. The pdf $g(t,\xi)$ may be calculated from the cumulative distribution function of $U(\xi)$ and its ``inverse'' process $S(t)= {\rm inf}\{ \xi: U(\xi)>t \}$ and is uniquely determined by the pdf of $U(\xi)$. In Ref. \cite{RSchilling10} it is shown that if the characteristic exponent $\widehat{\Psi}(s)$ is the completely Bernstein function (CBF), see Appendix \ref{a4}, then there exists its associated partner function $\widehat{\Phi}(s) = s/\widehat{\Psi}(s)$ which also is  CBF. The pair of $\widehat{\Psi}(s)$ and $\widehat{\Phi}(s)$ satisfies the relation $\widehat{\Psi}(s)\widehat{\Phi}(s)=s$ which is called the Sonine property, mathematical condition which enables reformulation of integral equations in terms integro-differential equations and \textit{vice versa} \cite{AHanyga20}. This unexpected duality has deeply meaningful consequences which have been noticed and discussed elsewhere \cite{AStanislavsky21,KGorska21}. Here they are only briefly mentioned in Sec. \ref{sec4}.  

\section{Spectral function for the KWW pattern}\label{sec2}

According to our best knowledge the analytic expression for the spectral function of KWW model was found in Refs. \cite{RHilfer02, RHilfer02a} and is not quoted elsewhere. Numerically it was calculated  in \cite{FAlvarez93} employing Eq. \eqref{15/07-1}. Calculations presented in Refs. \cite{RHilfer02, RHilfer02a} lead to representation of the KWW spectral in terms of the Fox $H$ function: 
\begin{equation}\label{27/07-1}
\widehat{\phi}_{KWW}(s) = 1 - H^{1, 1}_{1, 1}\left((s \tau)^{\alpha}\Big| {(1,1) \atop (1, \alpha)} \right), \quad \alpha\in(0, 1].
\end{equation}
According to their definition the Fox $H$ functions $H^{m, n}_{p, q}\big(z| {[a_{p}, A_{p}] \atop [b_{q}, B_{q}]}\big)$ are given by contour integrals of the Mellin-Barnes type (cf. Appendix \ref{a1}) where the upper list of parameters shortly denoted as $[a_{p}, A_{p}]$ means $(a_{1}, A_{1}), \ldots (a_{p}, A_{p})$. Similarly, the lower list of parameters shortly denoted as $[b_{q}, B_{q}]$ means $(b_{1}, B_{1}) \ldots (b_{q}, B_{q})$. Applying Eq. \eqref{31/08-1} to the Fox $H$ function of Eq. \eqref{27/07-1} we get
\begin{equation}\label{27/07-2}
H^{1, 1}_{1, 1}\left((s \tau)^{\alpha}\Big| {(1,1) \atop (1, \alpha)} \right) = \int_{L} \Gamma(1 + \alpha \xi) \Gamma(-\xi) (s\tau)^{-\alpha \xi} \frac{\D\xi}{2\pi\!\I},
\end{equation}
where the contour $L$ omits the poles of $\Gamma(1+\alpha\xi)$ and $\Gamma(-\xi)$. We remark that according to Eq. \eqref{15/07-1} in the Laplace space the Fox $H$ function of Eq. \eqref{27/07-1} equals to $s\, \widehat{n}_{KWW}(s)$. 

The Fox $H$ function is complicated mathematical object difficult to apply in practice. Except of its general properties only a little information useful in calculations can be found in standard compendia dealing with special functions \cite{Erdelyi53,Ryzhik00,APPrudnikov-v3}. Also it is not implemented in the computer algebra systems like Mathematica and Maple. All this makes calculations involving the Fox $H$ function difficult, time consuming and hard to be verified. To avoid this trouble we have found a way how to express $\widehat{\phi}_{KWW}(\I\!\omega)$ in terms of special functions which are analytically and numerically much more tractable. The solution goes as follows: begin with the observation that in any numerical calculation we are always restricted to using rational numbers such that the parameter $\alpha$ in Eq. \eqref{27/07-2} may be put equal to $l/k$. In such a case  the Fox $H$ in  Eq. \eqref{27/07-2} can be expressed in terms of the Meijer $G$ function $G^{m, n}_{p, q}\big(z| {(a_{p}) \atop (b_{q})}\big)$ (see Appendix \ref{a1}). Setting $\xi/k = -u$ in Eq. \eqref{27/07-2} we  rewrite the latter as
\begin{align}\label{27/07-3}
\begin{split}
& 1 - \widehat{\phi}_{KWW}(s) = k \int_{L_{u}} (s\tau)^{l u}\, \Gamma(1 - l u) \Gamma(k u) \D u/(2\pi\!\I) \\
& = \frac{\sqrt{l k}}{(2\pi)^{(l + k)/2 - 1}} \int_{L_{u}} \Big[\ulamek{l^{l}}{k^{k} (s\tau)^{l}}\Big]^{-u}\;  \prod_{i=0}^{k-1}\Gamma\big(\ulamek{i}{k} + u\big)\, \prod_{i=0}^{l-1} \\& \times \Gamma\big(\ulamek{1 + i}{l} - u\big)\, \frac{\D u}{2\pi\!\I} \\
& = \frac{\sqrt{l k}}{(2\pi)^{(l + k)/2 - 1}} G^{k, l}_{l, k}\Big(\ulamek{l^{l}}{k^{k} (s \tau)^{l}}\Big| {\Delta(l, 0) \atop \Delta(k, 0)}\Big),
\end{split}
\end{align}
where we have used the Gauss multiplication formula, i.e. $\Gamma(n z) = (2\pi)^{(1-n)/2} n^{nz - 1/2} \prod_{i=0}^{n-1} \Gamma(z + i/n)$, applied to $\Gamma(1-lu)$ and $\Gamma(k u)$. The upper and lower list of parameters are denoted as $\Delta(l, 0)$ and $\Delta(k, 0)$, respectively, where $\Delta(n, a)$ is a sequence of numbers $a/n, (a-1)/n, \ldots, (a+n-1)/n$. For  $l\leq k$, the Meijer $G$ function in \eqref{27/07-3} can be expressed as the finite sum of the generalized hypergeometric functions (see Appendix \ref{a1}) by using \cite[Eq. (8.2.2.3)]{APPrudnikov-v3}. 
\begin{align}\label{28/07-1}
1 - \widehat{\phi}_{KWW}(s) & = \sum_{j=0}^{k-1} \frac{(-1)^{j}}{j!}\, \frac{\Gamma(1 + \ulamek{l}{k}j)}{(s \tau)^{l j/k} } \nonumber\\ & \times {_{1+l}F_{k}}\left({1, \Delta(l, 1 + \frac{l}{k} j) \atop \Delta(k, 1+ j)};  \frac{(-1)^{k} l^{l}}{k^{k} (s\tau)^{l}}\right) \\ \label{28/07-1a}
& = \sum_{r\geq 0} \frac{(-1)^{r}}{r!} \, \frac{\Gamma(1 + l r/k)}{(s\tau)^{l r/k}}.
\end{align}
For passing between Eqs. \eqref{28/07-1} and \eqref{28/07-1a} we have used the series definition of generalized hypergeometric function ${_{p}F_{q}}$ and the formula in which the sum of $a_{r}$ is split into $k$ sums with the term $a_{k r}, a_{kr+1}, \ldots a_{kr +k-1}$, namely $\sum_{r\geq 0} a_{r} = \sum_{r \geq 0}\sum_{j=0}^{k-1} a_{kr + j}$. Note  that the result Eq. \eqref{28/07-1} we can get with the help of \cite[Eq. (2.3.2.13)]{APPrudnikov-v1}.
 
\section{Comparison of characteristic exponents}\label{sec4}

According to \cite{AStanislavsky15,AStanislavsky19} the characteristic (Laplace or L\'{e}vy) exponent may be retrieved from the knowledge the spectral function $\widehat{\phi}(s)$: 
\begin{equation}\label{26/07-4}
\widehat{\Psi}(s) = \frac{1 - \widehat{\phi}(s)}{\widehat{\phi}(s)}.
\end{equation}
{In the remaining part of the paper we assume $B(\tau) = 1$ such that all dependence from $\tau$ is shifted to the spectral functions $\widehat{\phi}(s)$}. The spectral functions for the CC, HN and JWS relaxation models are listed in Eq. \eqref{25/07-2} while for the KWW model it is given by Eq. \eqref{27/07-3} and/or Eq. \eqref{28/07-1}. Asymptotic behavior of all {considered} spectral functions for small and large frequencies confirms the experimentally established Jonscher's universal relaxation law (Jonscher's URL) \cite{AKJonscher92}. {From Refs. \cite{RHilfer02, RHilfer02a} or, independently, taking the Laplace transform of \cite[Eqs. (3.4), (3.29), and (3.45)]{RGarrappa16}} we get
\begin{multline}\label{30/07-1}
\hat{\phi}_{A; CC}(s) \sim \left\{
\begin{array}{c c}
(s\tau)^{-\alpha}, & s\tau \gg 1 \\
1-(s\tau)^{\alpha}, & s\tau \ll 1
\end{array}
\right., \\ 
\hat{\phi}_{A; HN}(s) \sim \left\{
\begin{array}{c c}
(s\tau)^{-\alpha\beta}, & s\tau \gg 1 \\
1-\beta\,(s\tau)^{\alpha}, & s\tau \ll 1
\end{array}
\right.,\\ \text{and} \qquad
\hat{\phi}_{A; JWS}(s) \sim \left\{
\begin{array}{c c}
\beta\,(s\tau)^{-\alpha}, & s\tau \gg 1 \\
1-(s\tau)^{\alpha\beta}, & s\tau \ll 1
\end{array}
\right.,
\end{multline}
where the parameters $\alpha$ and $\beta$ belong to the range $(0, 1]$. We put reader's attention that contrary to the CC model the asymptotics of the HN and JWS spectral functions {is governed by two different exponentials  - for the CC model it is only $\alpha$ while for the HN and JWS models $\alpha$ and $\alpha\beta=\gamma\leq\alpha\leq$. } 
%\footnote{{In what follows we will use the index $A$ to point out that we consider the asymptotics of given spectral function and/or characteristic functions.}}. 
This suggests to consider the CC and HN/JWS cases separately. 

To compare the characteristic exponents relevant for the above presented models we choose the KWW spectral function as the reference.  In Refs. \cite{RHilfer02, RHilfer02a} it was shown  that 
\begin{equation}\label{5/11-1}
\hat{\phi}_{A; KWW}(s) \sim \Gamma(1+\alpha)(s\tau)^{-\alpha}, \quad \text{for} \quad s\tau \gg 1,
\end{equation}
which flows out also from the series form of $\widehat{\phi}_{KWW}(s)$ given by Eq. \eqref{28/07-1a}.

\begin{figure}[h!]
\centering
\includegraphics[scale=0.2]{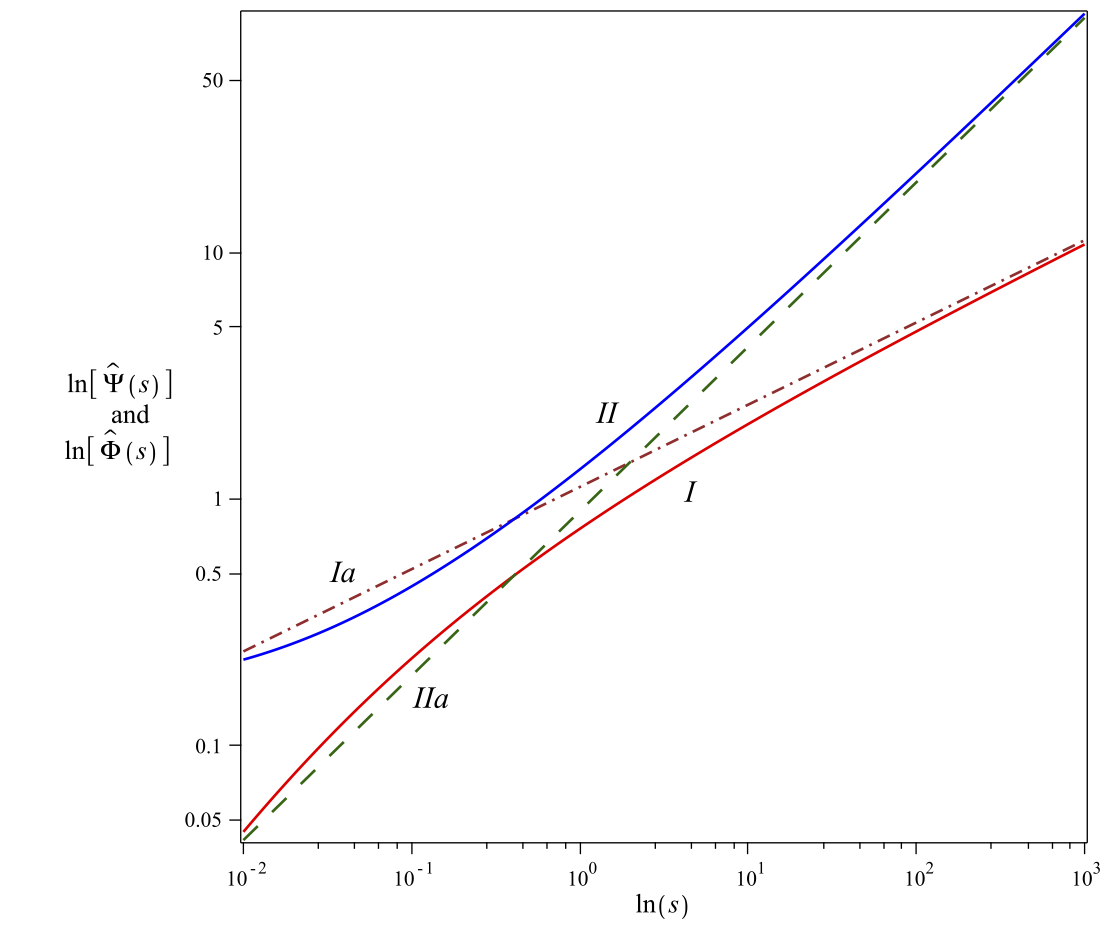}
\caption{\label{fig1} Logarithmic plot presents the comparison between characteristic functions $\widehat{\Psi}_{KWW}(s)$ (red solid curve no. {\em I}) and $\widehat{\Psi}_{CC}(s)$ (brown dot-dashed curve no. {\em Ia}) as well as between partner functions $\widehat{\Phi}_{KWW}(s)$ (blue solid curve no.{ \em II}) and $\widehat{\Phi}_{CC}(s)$ (green dashed curve no. {\em IIa}) for $\alpha = 1/3$ and $\tau = 1$. The characteristic exponents $\widehat{\Psi}_{KWW}(s)$ and $\widehat{\Phi}_{KWW}(s)$ have been calculated using Eq. \eqref{27/07-3} whereas $\widehat{\Psi}_{CC}(s)$ and $\widehat{\Phi}_{CC}(s)$ are given by Eqs. \eqref{2/08-2}. {The latter ones are, respectively, multiplied and divided by $[\Gamma(4/3)]^{-1}$.}}
\end{figure}

\smallskip
\noindent
{\bf (a)} The characteristic exponents for CC relaxation model are given by the power-law functions
\begin{equation}\label{2/08-2}
\widehat{\Psi}_{CC}(s) = (s\tau)^{\alpha} \qquad \text{and} \qquad \widehat{\Phi}_{CC}(s) = (s\tau)^{1-\alpha}
\end{equation}
{which differ from the asymptotic behavior of $\widehat{\Psi}_{KWW}(s)$ and $\widehat{\Phi}_{KWW}(s)$
\begin{multline}\label{12/08-1}
\widehat{\Psi}_{A; KWW}(s) \sim (s\tau)^{\alpha}/\Gamma(1+\alpha) \quad \text{and} \\ \widehat{\Phi}_{A; KWW}(s) \sim \Gamma(1+\alpha) (s\tau)^{1-\alpha}, \qquad s\tau \gg 1
\end{multline}
only by the factor $[\Gamma(1+\alpha)]^{-1}$. Thus, rescaling Eq. \eqref{2/08-2} we expect the asymptotic agreement with the characteristic functions of the KWW model calculated using Eqs.\eqref{27/07-3} or \eqref{28/07-1}.} The comparison of $\widehat{\Psi}_{KWW}(s)$ with $\widehat{\Psi}_{CC}(s)$ as well as $\widehat{\Phi}_{KWW}(s)$ with $\widehat{\Phi}_{CC}(s)$ are presented in Fig. \ref{fig1} where plots are made for $\alpha = 1/3$. It is seen that the characteristic exponents $\widehat{\Psi}_{CC}(s)$ and $\widehat{\Phi}_{CC}(s)$ match $\widehat{\Psi}_{KWW}(s)$ and $\widehat{\Phi}_{KWW}(s)$ for large $s$. {In the opposite  case, i.e.} for small $s$, $\widehat{\Psi}_{CC}(s)$ agrees with $\widehat{\Phi}_{KWW}(s)$ much better than $\widehat{\Psi}_{KWW}(s)$. Analogical observation can be made for $\widehat{\Phi}_{CC}(s)$ which reconstructs $\widehat{\Psi}_{KWW}(s)$ better than $\widehat{\Phi}_{KWW}(s)$. 
{We should also observe that $\widehat{\Psi}_{CC}(s)$ and $\widehat{\Phi}_{CC}(s)$ as well as $\widehat{\Psi}_{KWW}(s)$ and $\widehat{\Phi}_{KWW}(s)$ are CBFs and by construction satisfy the Sonine condition.}

\begin{figure}[h!]
\centering
\includegraphics[scale=0.22]{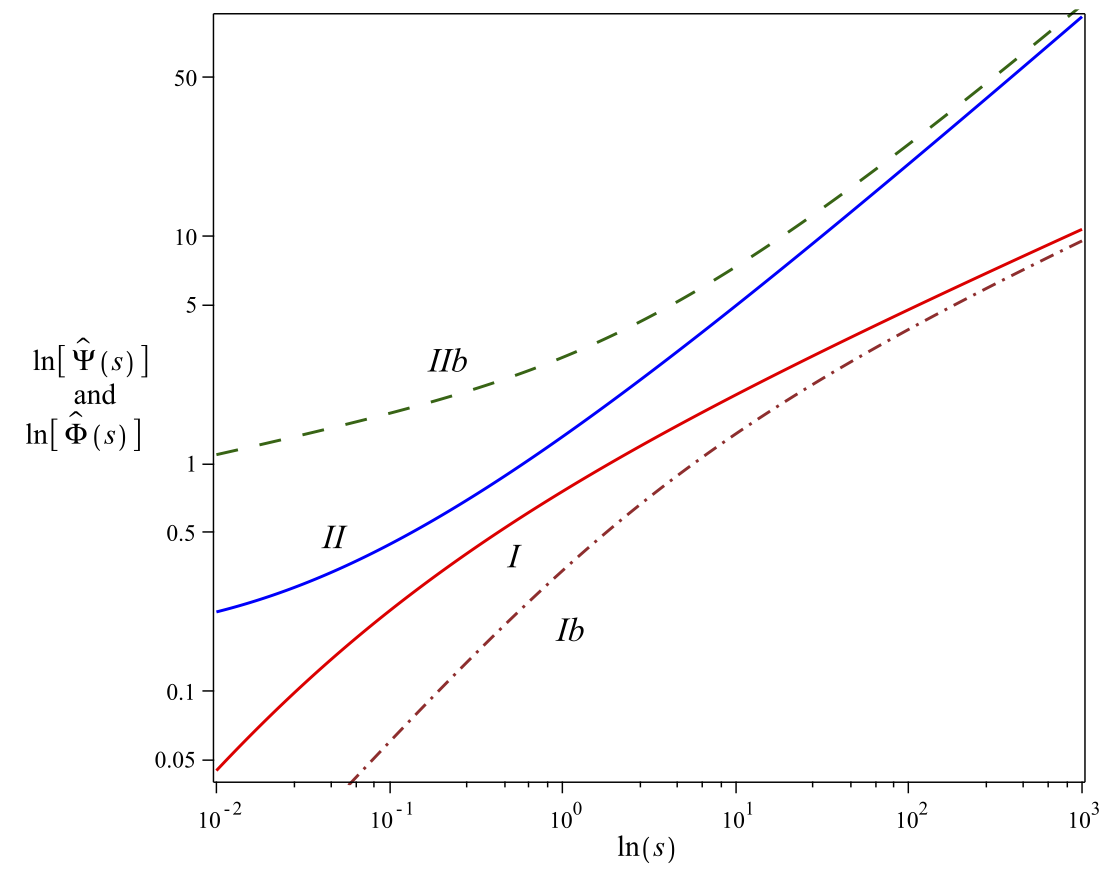}
\caption{\label{fig2} Logarithmic plot presents the comparison between characteristic exponents $\widehat{\Psi}_{KWW}(s)$ (red solid curve no. {\em I}) and $\widehat{\Psi}_{HN}(s)$ (brown dot-dashed curve no. {\em Ib}) as well as between partner functions $\widehat{\Phi}_{KWW}(s)$ (blue solid curve no. {\em II}) and $\widehat{\Phi}_{HN}(s)$ (green dashed curve no. {\em IIb}). $\widehat{\Psi}_{KWW}(s)$ and $\widehat{\Phi}_{KWW}(s)$ are calculated with the help of Eq. \eqref{27/07-3} where we use $\alpha_{KWW} = 1/3$, $\tau = 1$. The characteristic exponents $\widehat{\Psi}_{HN}(s)$ and $\widehat{\Phi}_{HN}(s)$ are given by Eqs. \eqref{3/08-5} where $\alpha_{HN} = 5/6$ and $\beta_{HN} = 2/5$. {The comparison between the characteristic functions is made with the factor $ [\Gamma(1+5/6)]^{-1}$ which is multiplied by $\widehat{\Psi}_{HN}(s)$ and divided by $\widehat{\Phi}_{HN}(s)$.} }
\end{figure}

\smallskip
\noindent
{\bf (b)} In case of the HN relaxation we have 
\begin{multline}\label{3/08-5}
\widehat{\Psi}_{HN}(s) = \{[1 + (s\tau)^{\alpha}]^{\beta} - 1\} \qquad \text{and} \\ \widehat{\Phi}_{HN}(s) = s\{[1 + (s\tau)^{\alpha}]^{\beta} - 1\}^{-1}.
\end{multline}
For large $s$ the leading asymptotic term of $\widehat{\Psi}_{HN}(s)$ is $(s\tau)^{\alpha\beta}$. For small $s$ te relevant asymptotics is got if we rewrite $\widehat{\Psi}_{HN}(s)$ as the series $\sum_{r\geq 0} \Gamma(1+\beta) (s\tau)^{{\alpha r}}/[r! \Gamma(1+\beta - r)] - 1$ whose first two terms (i.e. the terms with $r = 0$ and $r = 1$) give the asymptotics of $\widehat{\Psi}_{HN}(s)$ proportional to $\beta(s\tau)^{\alpha}$. Gathered together the asymptotic behavior of $\widehat{\Psi}_{HN}(s)$ reads 
\begin{multline}\label{3/08-1}
\widehat{\Psi}_{A; HN}(s) \sim (s\tau)^{\alpha\beta}, \quad s\tau \gg 1, \qquad \text{and} \\ \widehat{\Psi}_{A; HN}(s) \propto \beta (s\tau)^{\alpha}, \quad s\tau \ll 1
\end{multline}
which determine the asymptotics of $\widehat{\Phi}_{HN}(s)$
\begin{multline}\label{3/08-1a}
\widehat{\Phi}_{A; HN}(s) \sim (s\tau)^{1-\alpha}/\beta, \quad s\tau \gg 1, \qquad \text{and} \\ \widehat{\Phi}_{A; HN}(s) \sim (s\tau)^{1-\alpha\beta}, \quad s\tau \ll 1.
\end{multline}
{As in the previous example also here $\widehat{\Psi}_{A; HN}(s)$ and $\widehat{\Phi}_{A; HN}(s)$ are CBFs for $\alpha, \beta \in(0, 1]$.} The power-law asymptotics given by Eq. \eqref{3/08-1} for $s\tau\gg 1$ shows that in order to match $\widehat{\Psi}_{KWW}(s)$ the relations of exponentials $\alpha_{HN}\beta_{HN} = \alpha_{KWW}$ has to be satisfied. It means that $\alpha_{HN}$ may be chosen arbitrarily if simultaneously $\beta_{HN}=\alpha_{KWW}/\alpha_{HN}$. Thus the small $s$ asymptotics of $\widehat{\Psi}_{A; HN}(s)$ becomes incompatible with the asymptotics of $\widehat{\Psi}_{A; KWW}(s)$ and matches it only for $\beta=1$ which is the condition reducing the HN pattern to the CC one.  {Fig. \ref{fig2},  with $\alpha_{KWW} = 1/3$, $\alpha_{HN} = 5/6$, and $\beta_{HN} = 2/5$, shows that for large $\tau s$ $\widehat{\Psi}_{HN}(s)$ and $\widehat{\Phi}_{HN}(s)$ fit well $\widehat{\Psi}_{KWW}(s)$ and $\widehat{\Phi}_{KWW}(s)$, respectively, but the matching breaks down for small $s\tau$.} 

\begin{figure}[h!]
\centering
\includegraphics[scale=0.22]{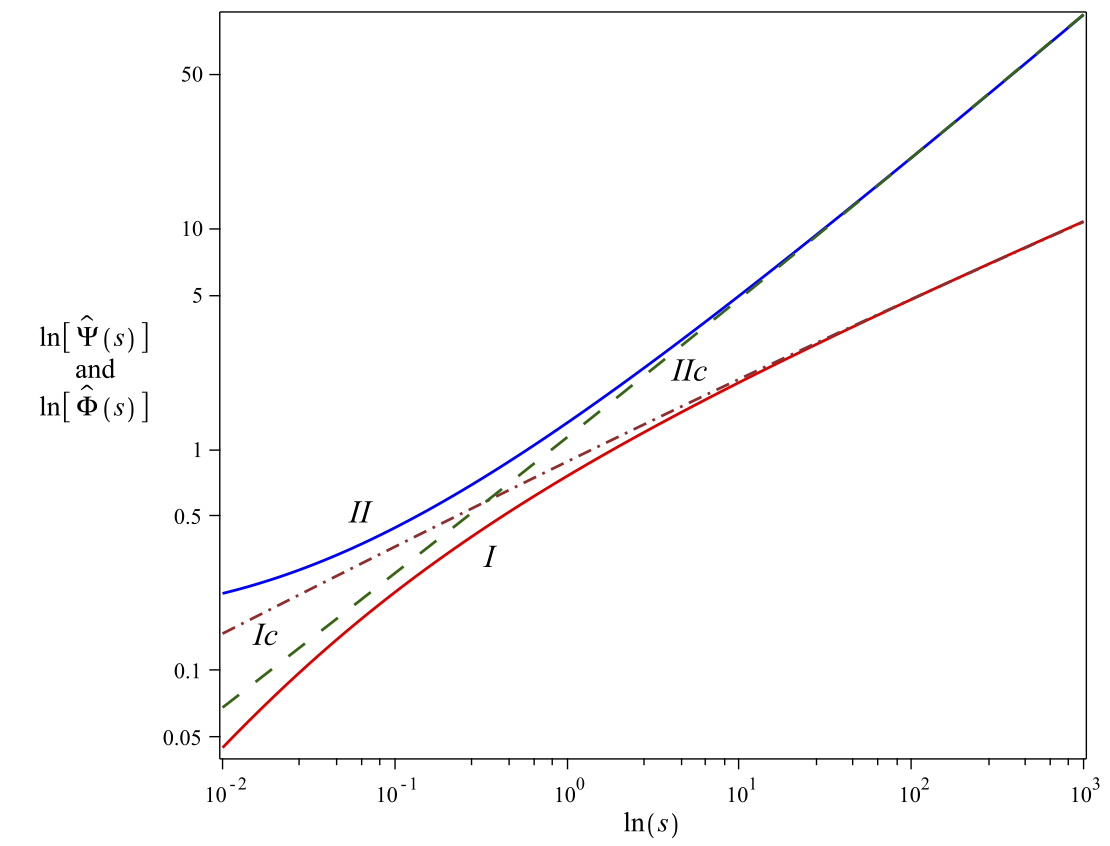}
\caption{\label{fig3} Logarithmic plot presents the comparison between $\widehat{\Psi}_{KWW}(s)$ (red solid curve no. {\em I}) and $\widehat{\Psi}_{JWS}(s)$ (brown dot-dashed curve no. {\em Ic}) as well as between $\widehat{\Phi}_{KWW}(s)$ (blue solid curve no. {\em II}) and $\widehat{\Phi}_{JWS}(s)$ (green dashed curve no. {\em IIc}). $\widehat{\Psi}_{KWW}(s)$ and $\widehat{\Phi}_{KWW}(s)$ are calculated with the help of Eq. \eqref{27/07-3} where we use $\alpha_{KWW} = 1/3$, $\tau = 1$. The characteristic exponents $\widehat{\Psi}_{JWS}(s)$ and $\widehat{\Phi}_{JWS}(s)$ are given by Eqs. \eqref{3/08-5} where $\alpha_{JWS} = 2/5$, $\beta_{JWS} = 5/6$, and {are, respectively, multiplied and divided by $[\Gamma(1 + 2/5)]^{-1}$.}}
\end{figure}

\smallskip
\noindent
{\bf (c)} The characteristic exponents of the JWS model are equal to 
\begin{multline}\label{4/08-1}
\widehat{\Psi}_{JWS}(s) = \{[1 + (s\tau)^{-\alpha}]^{\beta} - 1\}^{-1} \quad \text{and} \\ \widehat{\Phi}_{JWS}(s) = s \{[1 + (s\tau)^{-\alpha}]^{\beta} - 1\}
\end{multline}
Their asymptotics read
\begin{align}\label{4/08-3}
&\widehat{\Psi}_{A; JWS}(s) \sim \frac{s^{\alpha}}{\beta}, \,\, s\tau \gg 1; \quad \widehat{\Psi}_{A; JWS}(s) \sim (s\tau)^{\alpha\beta}, \,\,  s\tau \ll 1, \\ \label{4/08-3a}
&\widehat{\Phi}_{A; JWS}(s) \sim s^{1-\alpha\beta}, \,\,  s\tau \gg 1; \nonumber\\&\qquad\qquad\qquad\quad \widehat{\Phi}_{A; JWS}(s) \sim \frac{\beta}{(s\tau)^{\alpha-1}}, \,\,  s\tau \ll 1.
\end{align}
{As in the previous case also here the asymptotics' presented by Eq. \eqref{4/08-3} are given by CBFs for $\alpha, \beta \in(0, 1]$.} The comparison between $\widehat{\Psi}_{KWW}(s)$ and $\widehat{\Psi}_{JWS}(s)$ as well as between $\widehat{\Phi}_{KWW}(s)$ and $\widehat{\Phi}_{JWS}(s)$ is shown in Fig. \ref{fig3}. It is seen that for large $s$ $\widehat{\Psi}_{JWS}(s)$ and $\widehat{\Phi}_{JWS}(s)$ match $\widehat{\Psi}_{KWW}(s)$ and $\widehat{\Phi}_{KWW}(s)$ faster than {for} the CC and HN models. Nevertheless the matchings for small $s$ remain disappointing although at the first glance they seem to be more acceptable than those resulting from the CC and HN models. This, however, may be treated as an artefact coming from the choice of parameters.  

We complete this section with two remarks:\\
\smallskip
\noindent
1. The leading order of large $s$, i.e. short $t$, asymptotics of all relaxation patterns being considered matches the KWW function.  \\
\smallskip
\noindent
2. In Figs. \ref{fig1}-\ref{fig3} the curves labelled by {\em I} and {\em II} show the behavior of various  $\widehat{\Psi}(s)$ and $\widehat{\Phi}(s)$. Unidexed labels {\em I} and {\em II} characterize plots obtained for  the KWW model if $\alpha = 1/3$. The labels {\em I} and {\em II} indexed with subscripts $a$, $b$, and $c$ distinguish non-Debye models: $a$ is for the CC, $b$ is for the HN, and $c$ is for the JWS.

\section{{The Mittag-Leffler family: comparison of useful properties}}\label{sec5}

\subsection{An interlude: a few mathematical tools}

\subsubsection{{The Efross theorem as an integral decomposition}} 

The Efross theorem \cite{AEfross35, LWlodarski52, UGraf04, KGorska12a, AApelblat21, KGorska21b}  generalizes the Borel convolution theorem for the Laplace transform. According to it, for $\widehat{G}(s)$ and $\widehat{q}(s)$ being analytic functions,  we have
\begin{align}\label{6/08-3}
\begin{split}
\mathscr{L}^{-1}&\left[\widehat{G}(s)\, \widehat{h}_{1}(x,\widehat{q}(s)); t\right] \\ &= \int_{0}^{\infty} \mathscr{L}^{-1}[\widehat{h}_{1}(x, s); \xi]\, \mathscr{L}^{-1}[\widehat{G}(s) \E^{-\xi\, \widehat{q}(s)}; t] \D\xi\\
& = \int_{0}^{\infty} h_{1}(x, \xi) h_{2}(\xi, t) \D\xi .
\end{split}
\end{align}
in which one immediately recognizes the structure of integral decomposition. In the probabilistic language, if $h_{1}(x, \xi)$ and $ h_{2}(\xi, t)$ are independent probability distributions, Eq. \eqref{6/08-3} expresses the Bayes theorem and thus may be treated as a joint probability distributions. Namely this identification is made when stochastic methods are applied to the relaxation theory \cite{HCFogedby94}. Within this approach the non-negative random variable $\xi$ is interpreted as an "internal" or operational time which governs the evolution of the function $h_{1}(x, \xi) = \mathscr{L}^{-1}[\widehat{h}_{1}(x, s); \xi]$ bearing the name of the parent process. The second component of the integral decomposition Eq. \eqref{6/08-3} $h_{2}(\xi, t) = \mathscr{L}^{-1}[\widehat{G}(s) \E^{-\xi\, \widehat{q}(s)}; t]$ describes the mutual dependence of operational $\xi$ and physical $t$ times. Unlike regularly clocked physical time $t$ the internal time $\xi$ has the nature of a \textit {c\`{a}dl\`{a}g} (left continuous right limited) nonnegative and non-decreasing stochastic process. According to classification proposed in Ref. \cite{IMSokolov02} and recently reconsidered in Refs. \cite{ AChechkin21, KGorska21b} both functions $h_{i}$'s, $i = 1, 2$, can be either the ``safe'' or ``dangerous'' probability densities (PDFs). Sufficient condition to be the "safe" PDF is infinite divisibility, if it is not the case we may deal with an example of  ``dangerous'' PDF. The working criterion to distinguish the ``safe'' and ``dangerous'' cases is their adherence to the class of Bernstein function. This guarantees that the features characterizing the ``safe'' PDFs, i.e. nonnegativity and infinite  divisibility, are satisfied (Appendix \ref{a3}). Remember also that $h_{2}(\xi, t)$ should be normalized. For $h_{2}(\xi, t)$ being the ``safe'' PDF we can say that it subordinates $h_{1}(x, \xi)$. In the opposite case, i.e. for $h_{2}(\xi, t)$ being the ``dangerous'' PDF, we name $h_{2}(\xi, t)$ and $h_{1}(x, \xi)$ the constituents of integral decomposition only.

\subsubsection{{Integral decompositions as subordinations}}

In the case of relaxation theory we know from Eq. \eqref{26/07-3}  that $h_{1}(x, \xi)$ is independent on $x$. It is equal to the Debye relaxation function, i.e., $h_{1}(\xi) \equiv n_{D}(\xi) = \exp(- B \xi)$. The latter is CBF as it is the nonnegative, normalized, and infinitely divisible with respect to $\xi$. Thus, $n_{D}(\xi)$ is the PDF of parent process. The function $h_{2}(\xi, t) \equiv h_{2, \Psi}(\xi, t)$ involves $\widehat{G}(s) = \widehat{\Psi}(s)/s$ as well as $\widehat{q}(s) = \widehat{\Psi}(s)$ and it is equal to $\mathscr{L}^{-1}\{[\widehat{\Psi}(s)/s] \exp[-\xi \widehat{\Psi}(s)]; t\}$. Normalization of $h_{2, \Psi}(\xi, t)$ in $\xi$ is fulfilled automatically. Because $\widehat{\Psi}(s)$ is the characteristic exponent, i.e. it belongs to the class of Bernstein functions, then using Appendix \ref{a3} we can show that $h_{2, \Psi}(\xi, t)$ is ``safe'' PDF and it subordinates the Debye relaxation process $h_{1}(\xi)$.  With the help of Efross theorem Eq. \eqref{6/08-3} $n(t)$ can be written as
\begin{align}\label{6/08-4}
\begin{split}
n(t) & = \int_{0}^{\infty} \mathscr{L}^{-1}[\widehat{n}_{D}(s); \xi]\, \mathscr{L}^{-1}\Big[\frac{\widehat{\Psi}(s)}{s} \E^{-\xi \widehat{\Psi}(s)}; t\Big] \D\xi \\
& = \mathscr{L}^{-1}\Big[\frac{\widehat{\Psi}(s)}{s}\, \widehat{n}_{D}\big(\widehat{\Psi}(s)\big); t\Big],
\end{split}
\end{align}
where $\widehat{n}_{D}(s) = \mathscr{L}[n_{D}(t); s] = (B + s)^{-1}$. For $\widehat{\Psi}(s)$ being CBF there exists associated CBF $\widehat{\Phi}(s)$ such that $\widehat{\Psi}(s) \widehat{\Phi}(s) = s$. Thus, Eq. \eqref{6/08-4} can be written down in its alternative form
\begin{align}\label{15/08-1}
\begin{split}
n(t) & = \mathscr{L}^{-1}\Big\{[\widehat{\Phi}(s)]^{-1}\, \widehat{n}_{D}\big(s/\widehat{\Phi}(s)\big); t\Big] \\
& = \int_{0}^{\infty} \mathscr{L}^{-1}[\widehat{n}_{D}(s); \xi]\, \mathscr{L}^{-1}\Big\{[\widehat{\Phi}(s)]^{-1}\E^{-\xi s/\widehat{\Phi}(s)}; t\Big\} \D\xi
\end{split}
\end{align}
with $h_{2}(\xi, t)~\equiv~h_{2,\Phi}(\xi, t)~=~\mathscr{L}^{-1}\{\exp[-\xi s/\widehat{\Phi}(s)]/\widehat{\Phi}(s); t\}$ subordinates the Debye relaxation process as well. This duality is not problematic - both characteristic functions  $\Psi(s)$ and $\Phi(s)$ lead to the same $n(t)$ if $\widehat{\Psi}(s) \widehat{\Phi}(s) = s$ which once more emphasizes the importance of the Sonine condition. Also, by virtue of considerations presented in Sec. \ref{sec1a} (cf. Eq. \eqref{26/07-3}), we know that $h_{2, \Psi}(\xi, t)$ and $h_{2, \Phi}(\xi, t)$ are always the "safe" PDFs.  Thus they may be used to subordinate the Debye relaxation as well as other "safe" distributions, e.g., the normal distribution, cf. Ref. \cite{AStanislavsky21}. We can also use another subordinators, e.g. $\mathscr{L}^{-1}\{s^{\alpha\gamma-\beta} \exp[-u s^{\alpha}]; t\}$, which subordinates the CD  relaxation model. The example of using two kinds of subordinators is presented in Ref. \cite{KGorska21b}. 

\subsubsection{{Subordinations as signposts leading to evolution equations}}

{The subordination approach allows one to find the evolution equations which govern the behavior of subordinated parent process \cite{IMSokolov02}; this we have named the relaxation function $n(t)$.} To obtain suitable equations we need two building blocks. The first of them is the standard evolution equation for ${n}_{D}(t)$, considered in the Laplace space with $s$ replaced by $\widehat{\Psi}(s)$. The second one is the relation $\widehat{n}_{D}(\,\widehat{\Psi}(s)) = s \widehat{n}(s)/\widehat{\Psi}(s)$ derived from Eq. \eqref{6/08-4}. 

The evolution equation (with respect to the time $t$) of $n_{D}(t)$ is well-known. It reads $\dot{n}_{D}(t) = - B\, n_{D}(t)$ and in the Laplace space {equals to}  
\begin{equation}\label{6/08-5}
s \widehat{n}_{D}(s) - n_{D}(0) = - B\, \widehat{n}_{D}(s)
\end{equation}
with the initial condition $n_{D}(0) = 1$. After replacing $s$ by $\widehat{\Psi}(s)$ we have
\begin{equation}\label{6/08-6}
\widehat{\Psi}(s)\, \widehat{n}_{D}\big(\,\widehat{\Psi}(s)\big) - 1 = - B\, \widehat{n}_{D}\big(\widehat{\Psi}(s)\big).
\end{equation}
Setting $\widehat{n}_{D}(\,\widehat{\Psi}(s)) = s \widehat{n}(s)/\widehat{\Psi}(s)$ and substituting it into Eq. \eqref{6/08-6} {we get}
\begin{equation}\label{6/08-7}
\widehat{n}(s) = s^{-1} - B\, [\widehat{\Psi}(s)]^{-1} \widehat{n}(s),
\end{equation}
which can be rewritten  as
\begin{equation}\label{30/08-1}
\frac{\widehat{\Psi}(s)}{s}[s\, \widehat{n}(s) - 1] = -B \widehat{n}(s).
\end{equation}
Taking the inverse Laplace transform of Eqs. \eqref{6/08-7} and \eqref{30/08-1} in which we use the Laplace convolution we get, respectively,
\begin{multline}\label{6/08-8}
n(t) = 1 - B \int_{0}^{t} M(t - u) n(u) \D u \quad \text{and} \\ \int_{0}^{t} k(t - u) \dot{n}(u) \D u = - B n(t),
\end{multline}
where
\begin{equation}\label{7/08-1}
M(t) = \mathscr{L}^{-1}[\widehat{M}(s); t] \quad \text{with} \quad \widehat{M}(s) = [\widehat{\Psi}(s)]^{-1},
\end{equation}
and
\begin{equation}\label{30/08-2}
k(t) = \mathscr{L}^{-1}[\widehat{k}(s); t] \quad \text{with} \quad \widehat{k}(s) = [\widehat{\Psi}(s)/s]^{-1} = [\widehat{\Phi}(s)]^{-1}.
\end{equation}
The functions $M(t)$ and $k(t)$ are nonnegative and interpreted as memory functions or kernels. Because $\widehat{\Psi}(s)$ is CBF then $\widehat{M}(s)$ is the Stieltjes function (SF) and $s/\widehat{\Psi}(s)$ is also CBF, see Appendix \ref{a4}. Denoting the latter as $\widehat{\Phi}(s)$ we learn that $\widehat{k}(s)$ is also SF \cite{KGorska20, KGorska21, KGorska21a}. Using the relation $\widehat{\Psi}(s) \widehat{\Phi}(s) = s$ shows that $\widehat{M}(s)\, \widehat{k}(s) = s^{-1}$ which means that the memory functions $M(t)$ and $k(t)$ fulfill  the Sonine equation
\begin{equation}\label{8/08-1}
\int_{0}^{t} M(t - u) k(u) \D u = \int_{0}^{t} M(u) k(t - u) \D u = 1.
\end{equation}
Eqs. \eqref{6/08-8} express the time smeared evolution, either of the relaxation function or its derivative. Detailed discussion of mutual relation between these equations has been presented in \cite{KGorska21, KGorska21a}. From the physical point of view the first equation present in the pair  \eqref{6/08-8} is known as the master equation and may be considered as general modeling of the memory dependent linear evolution scheme. Mathematically Eqs. \eqref{6/08-8} are both the Volterra type equations \cite{GGripenberg90} which shape and utility goes beyond much more popular fractional differential equations introduced in the framework of fractional calculus approach to the relaxation phenomena \cite{KGorska20, RGarrappa16}.

\subsection{{Examples}}

As noticed, Eqs. \eqref{6/08-4} and \eqref{15/08-1} yield the same results {as governed by equivalent Eqs. \eqref{6/08-8}.} Thus, the relaxation function $n(t)$ can be derived with the help either of Eqs. \eqref{6/08-4} or \eqref{15/08-1}. Without loss of generality we take Eq. \eqref{6/08-4}, the characteristic function $\widehat{\Psi}(s)$ presented in Sec. \ref{sec3} {and more convenient for our purposes this one of Eqs. \eqref{6/08-8} which describes the evolution of $n(t)$}. 

To derive $n(t)$ for any among the non-Debye relaxation models listed in Sec. \ref{sec3} we will use the formula 
\begin{equation}\label{15/08-3}
\mathscr{L}^{-1}[s^{\alpha\gamma - \beta}\E^{- u s^{\alpha}}; t] = \frac{\Gamma(\gamma)\, t^{\beta}}{\alpha\, u^{\gamma}} g_{\alpha, \beta}^{\gamma}(u, t)
\end{equation}
obtained from \cite[Eqs. (5) and (6)]{KGorska20a} with the help of the second formula in Eq. \eqref{30/07-5}. The function $g_{\alpha, \beta}^{\gamma}(u, t)$ is connected to its one variable version $g_{\alpha, \beta}^{\gamma}(x)$ through Eq. \eqref{30/07-6}. Eq. \eqref{15/08-3} generalizes the known formula $\mathscr{L}^{-1}[s^{\alpha-1} \E^{-u s^{\alpha}}; t] = t g_{\alpha}(u, t)/(\alpha u)$ appearing for the one-parameter Mittag-Leffer function \cite{HPollard48, KGorska12a, KGorska21b}. Indeed, Eq. \eqref{15/08-3} reduces to it for $\gamma = \beta = 1$. Having all this at  hands we are ready to find relaxation functions looked for.

\smallskip
\noindent
{\bf (i)} For the CC model, for which $\widehat{\Psi}_{CC}(s)$ is given by Eq. \eqref{2/08-2} multiplied by $B$, we have
\begin{align}\label{15/08-2}
\begin{split}
n_{CC}(t) & = \int_{0}^{\infty} \E^{-B u} \mathscr{L}^{-1} [B \tau^{\alpha} s^{\alpha -1} \E^{-u B \tau^{\alpha}s^{\alpha}}; t] \D u \\
& = \int_{0}^{\infty} \E^{-\tau^{-\alpha} \xi} \mathscr{L}^{-1}[s^{\alpha-1} \E^{-\xi s^{\alpha}}; t] \D\xi,
\end{split}
\end{align}
where we set $u B \tau^{\alpha} = \xi$. We say that $\mathscr{L}^{-1}[s^{\alpha-1} \E^{-\xi s^{\alpha}}; t]$ subordinates the Debye case. Using Eq. \eqref{15/08-3} for $\gamma = \beta = 1$ we get Eq. \eqref{30/07-3} with $a = \tau^{-\alpha}$ and $p = t$ and 
\begin{equation}\label{16/08-1}
n_{CC}(t) = E_{\alpha}[-(t/\tau)^{\alpha}],
\end{equation}
which is well-known relaxation function of the CC model \cite{RHilfer02, RHilfer02a, RGarrappa16, KWeron96}.  Asymptotics of $n_{CC}(t)$ for $t$ being smaller or larger than $\tau$ can be obtained from the first terms of the series representations given in Refs. \cite{RHilfer02, RHilfer02a, RGarrappa16}. The results  read
\begin{multline}\label{18/08-1}
n_{A; CC}(t) \sim 1 - \frac{(t/\tau)^{\alpha}}{\Gamma(1+\alpha)}, \quad t\ll \tau, \quad \text{and} \\ n_{A; CC}(t) \sim \frac{(t/\tau)^{-\alpha}}{\Gamma(1-\alpha)}, \quad t \gg \tau.
\end{multline}
{The short time asymptotics given by the first equation in Eqs. \eqref{18/08-1} constitutes also the first two terms of the stretched exponential $\exp[-t^{\alpha}/\Gamma(1+\alpha)]$. Such approximation was proposed in Ref. \cite{FMainardi14} and it offers good results for low values of $\alpha$ at sufficiently short times.}

{The evolution equations derived from} Eqs.  \eqref{6/08-8} read
\begin{multline}\label{9/08-1}
n_{CC}(t) = 1 -\tau^{-1} (I^{\alpha} n_{CC})(t) \quad \text{and} \\ ({^{c}D^{\alpha} n_{CC}})(t) = - \tau^{\alpha} n_{CC}(t),
\end{multline}
where {the symbol $(I^{1-\alpha} \frac{\D}{\D x} f)(x)$ denotes the fractional integral defined in Appendix \ref{a5} for $\alpha\in(0, 1)$ while ${^{c}D^{\alpha}}$ is the fractional (Caputo) derivative operator.} Eqs. \eqref{9/08-1} is equivalent to \cite[Eq. (3.7.43)]{RGorenflo14} and  \cite[Eq. (3.10)]{RGarrappa16}. 

\smallskip
\noindent
{\bf (ii)} Our next example is the HN relaxation function. We begin with the subordination {approach} which involves the Debye relaxation and $\mathscr{L}^{-1}\{\widehat{\Psi}(s) \exp[-\xi\widehat{\Psi}(s)]/s; t\}$. Substituting Eq. \eqref{3/08-5} {multiplied by $B$} into Eq. \eqref{6/08-4} we get
\begin{multline}\label{2/09-1}
n_{HN}(t)  = B \int_{0}^{\infty} \mathscr{L}^{-1}\{Bs^{-1}[\tau^{\alpha\beta}(\tau^{-\alpha} + s^{\alpha})^{\beta} - 1] \\ \times \E^{-\xi B \tau^{\alpha\beta}(\tau^{-\alpha} + s^{\alpha})^{\beta}}; t\} \D\xi,
\end{multline}
{where $n_{D}(\xi)$ is cancelled by $\exp(-B\xi)$ coming from $\widehat{\Psi}_{HN}(s)$.} Next, we apply once more the Efross theorem to the inverse Laplace transform in Eq. \eqref{2/09-1}, this time with $\widehat{G}(s) = s^{-1}$ and $\widehat{q}(s) = \tau^{-\alpha} + s^{\alpha}$ put in. Thus, we can express Eq. \eqref{2/09-1} as
\begin{align}\label{2/09-2}
\begin{split}
n_{HN}(t) & = B \int_{0}^{\infty} \left\{\int_{0}^{\infty} \mathscr{L}^{-1}[(\tau^{\alpha\beta} s^{\beta} - 1) \E^{-\xi B \tau^{\alpha\beta} s^{\beta}}; u] \right.\\ & \left. \times \E^{-u\tau^{-\alpha}} \mathscr{L}^{-1}[s^{-1} \E^{-u s^{\alpha}}; t] \D u\right\} \D\xi \\
& = B \int_{0}^{\infty} \mathscr{L}^{-1}\Big[(\tau^{\alpha\beta} s^{\beta} - 1) \int_{0}^{\infty} \E^{-\xi B \tau^{\alpha\beta} s^{\beta}} \D\xi; u\Big] \\ & \times \mathscr{L}^{-1}[s^{-1} \E^{-u (\tau^{-\alpha} + s^{\alpha})}; t] \D u \\
& = \int_{0}^{\infty} \E^{-u\tau^{-\alpha}} \mathscr{L}^{-1}[s^{-\beta}(s^{\beta} - \tau^{-\alpha\beta}); u] \\& \times \mathscr{L}^{-1}[s^{-1} \E^{-u s^{\alpha}}; t] \D u.
\end{split}
\end{align}
Because of $$\mathscr{L}^{-1}[s^{-\beta}(s^{\beta} - \tau^{-\alpha\beta}); u]~=~\delta(u)~-~\tau^{-\alpha\beta} u^{\beta-1}/\Gamma(\beta)$$ we rewrite Eq. \eqref{2/09-2} as
\begin{align}\label{16/08-3a}
n_{HN}(t) & =  1 - \tau^{-\alpha\beta} \int_{0}^{\infty} \E^{-\tau^{-\alpha} u} \frac{u^{\beta-1}}{\Gamma(\beta)}\, \mathscr{L}^{-1}[s^{-1} \E^{-u s^{\alpha}}; t] \D u \\ \label{16/08-3b}
& = 1 - (t/\tau)^{\alpha\beta} \int_{0}^{\infty} \E^{-\tau^{-\alpha} u}\, \frac{t}{\alpha u} g_{\alpha, 1 + \alpha\beta}^{\beta}(u, t) \D u.
\end{align} 
From Eq. \eqref{15/08-3} it comes out that  $\mathscr{L}^{-1}[s^{-1} \E^{-u s^{\alpha}}; t] = \Gamma(\beta) t^{1 + \alpha\beta}/(\alpha u^{\beta}) g_{\alpha, 1 + \alpha\beta}^{\beta}(u, t)$.  Then, with the help of Eq. \eqref{30/07-5}, we get
\begin{equation}\label{17/08-1}
 n_{HN}(t) = 1 - (t/\tau)^{\alpha\beta} E_{\alpha, 1 + \alpha\beta}^{\beta}[-(t/\tau)^{\alpha}],
\end{equation} 
which is usually obtained by employing $\widehat{\phi}_{HN}(\I\!\omega)$, Eqs. \eqref{25/08-1}, and \eqref{15/07-1} as it is presented in \cite{RHilfer02, RHilfer02a, RGarrappa16, KGorska18}. Notice that due to \cite[Eq. (3.13)]{RGarrappa16} we have that 
\begin{equation}\label{2/09-10}
\E^{-\tau^{-\alpha} u} \frac{u^{\beta-1}}{\Gamma(\beta)} = \tau^{\alpha\beta}\mathscr{L}^{-1}[(1 + s\tau^{\alpha})^{-\beta}; t],
\end{equation}
proportional to the spectral function of CD model which now generates the parent process. Hence, looking on Eqs. \eqref{16/08-3a} and \eqref{2/09-10} we can say that the CD relaxation together with $\mathscr{L}^{-1}[s^{-1}\exp(-u s^{\alpha}); t]$ are the constituents of the integral decomposition of $1-n_{HN}(t)$. From \cite{RHilfer02, RHilfer02a, RGarrappa16} we know that the short and long time power-law asymptotics of $n_{HN}(t)$ are equal to
\begin{multline}\label{18/08-2}
n_{A; HN}(t) \sim 1 - \frac{(t/\tau)^{\alpha\beta}}{\Gamma(1 + \alpha\beta)}, \quad t \ll \tau, \quad \text{and} \\ n_{A; HN}(t) \sim \frac{\beta\, (t/\tau)^{-\alpha}}{\Gamma(1-\alpha)}, \quad t \gg \tau.
\end{multline}

{The evolution equation is equal to \cite[Eqs. (3.40) and (3.50)]{RGarrappa16}}
\begin{equation}\label{9/08-2}
{^{\rm C}}(D^{\alpha} + \tau^{-\alpha})^{\beta} n_{HN}(t) = -\tau^{\alpha\beta},
\end{equation}
{where the pseudo--operator ${^{\rm C}}(D^{\alpha} + \tau^{-\alpha})^{\beta}$ is defined in Appendix \ref{a5}.}

\smallskip
\noindent
{\bf (iii)} Eq. \eqref{6/08-4} for the JWS model gives
\begin{multline}\label{2/09-3}
n_{JWS}(t) = B \int_{0}^{\infty} \mathscr{L}^{-1}\Big[\frac{s^{\alpha\beta - 1}}{(s^{\alpha} + \tau^{-\alpha})^{\beta} - s^{\alpha\beta}} \\ \times \E^{-\xi B \frac{(s^{\alpha} + \tau^{-\alpha})^{\beta}}{(s^{\alpha} + \tau^{-\alpha})^{\beta} - s^{\alpha\beta}}}; t\Big] \D\xi,
\end{multline}
which after using once again the Efross theorem where $\widehat{G}(s) = s^{\alpha\beta - 1}$ and $\widehat{q}(s) = s^{\alpha} + \tau^{-\alpha}$ can be represented as
\begin{multline*}%\label{2/09-4}
n_{JWS}(t) = B \\ \times \int_{0}^{\infty}\!\! \left\{\int_{0}^{\infty}\!\! \mathscr{L}^{-1}\Big[\frac{1}{s^{\beta} - (s -\tau^{-\alpha})^{\beta}} \E^{-\xi B \frac{s^{\beta}}{s^{\beta} - (s - \tau^{-\alpha})^{\beta}}}; u\Big] \right.\\ \left. \times \mathscr{L}^{-1}[s^{\alpha\beta - 1}\E^{-u (\tau^{-\alpha} + s^{\alpha})}; t]\D u\right\} \D\xi. 
\end{multline*}
Calculating the integral over $\xi$ (as it was done in the example {\bf (ii)}) we can simplify the first inverse Laplace transform. It enables us to write down the above equation in the form 
\begin{align}\label{2/09-5}
\begin{split}
n_{JWS}(t) & = \int_{0}^{\infty} \mathscr{L}^{-1}[s^{-\beta}; u] \E^{- u \tau^{-\alpha}} \mathscr{L}^{-1}[s^{\alpha\beta - 1}\E^{-u s^{\alpha}}; t]\D u \\
& = \int_{0}^{\infty} \E^{- u \tau^{-\alpha}} u^{\beta-1}/\Gamma(\beta) \mathscr{L}^{-1}[s^{\alpha\beta - 1}\E^{-u s^{\alpha}}; t]\D u \\
& = \int_{0}^{\infty} \E^{-\tau^{-\alpha}} \frac{t}{\alpha u} g_{\alpha, 1}^{\beta}(t, u) = E_{\alpha, 1}^{\beta}[-(t/\tau)^{\alpha}].
\end{split}
\end{align}
To show that  $\mathscr{L}^{-1}[s^{\alpha\beta - 1} \E^{-u s^{\alpha}}; t]=\Gamma(\beta) t/(\alpha u^{\beta}) g_{\alpha, 1}^{\beta}(u, t)$ we employ Eq. \eqref{15/08-3}. Next, using Eq. \eqref{2/09-10} we express $n_{JWS}(t)$ as
\begin{multline}\label{2/09-11}
n_{JWS}(t) = \int_{0}^{\infty} \mathscr{L}^{-1}[(\tau^{-\alpha} + s^{\alpha})^{-\beta}; u]\\ \times \mathscr{L}^{-1}[s^{\alpha\beta - 1} \E^{-u s^{\alpha}}; t] \D u,
\end{multline}
which means that to obtain the relaxation function of the JWS model we can use two kinds of subordination approaches. Using the different subordinators we can subordinate the Debye or CD process. Thus, the processes which lead to the JWS relaxation model can be obtained using the various approaches. The asymptotic behaviour of $n_{JWS}(t)$ can be given by
\begin{multline}\label{18/08-3}
n_{A; JWS}(t) \sim 1 - \frac{\beta\, (t/\tau)^{\alpha}}{\Gamma(1 + \alpha)}, \quad t \ll \tau, \quad \text{and} \\ n_{A; JWS}(t) \sim \frac{\beta\, (t/\tau)^{-\alpha\beta}}{\Gamma(1-\alpha\beta)}, \quad t \gg \tau.
\end{multline}

{Its evolution equation reads
\begin{equation}\label{9/08-3}
(D^{\alpha} + \tau^{-\alpha})^{\beta} n_{JWS}(t) = \frac{t^{\alpha\beta}}{\Gamma(1-\alpha\beta)}.
\end{equation}
Appendix \ref{a5} contains the definition of the pseudo-operator $(D^{\alpha} + \tau^{-\alpha})^{\beta}$. Eq. \eqref{9/08-3} has been discussed in \cite[Eq. (4.3)]{AStanislavsky16} and justified by under- and overshooting subordination technique applied for the anomalous diffusion.}

Analyzing the above examples we see that the integral decomposition Eq. \eqref{6/08-3} can be interpreted as an alternative form of Eq. \eqref{15/07-1} obtained {from Eq. \eqref{26/07-3}} by employing Eq. \eqref{26/07-4}. {It provides us also missing long time asymptotics of the KWW relaxation model Eq. \eqref{25/07-1}.  The relevant asymptotic behavior for short and long times is  
\begin{multline}\label{18/08-5}
n_{KWW}(t) \sim 1 - (t/\tau)^{\alpha}, \quad t \ll \tau, \quad \text{and} \\ n_{KWW}(t) \sim \exp[-(t/\tau)^{\alpha}], \quad t \gg \tau,
\end{multline}
obtained in Refs. \cite{RHilfer02, RHilfer02a} .} 

Looking at the asymptotics of CC and JWS relaxation functions we see that for short times they have similar power-law behavior as $n_{KWW}(t)$ albeit they differ by a constant. The HN relaxation function differs more significantly - the exponential involves both parameters and to agree the asymptotics we have to put restrictive condition $\alpha_{HN}\beta_{HN} = \alpha_{KWW}$. Analogous, but reverse, situation we met when want to agree the asymptotics for large $t$. This leads to the conclusion going beyond the short time asymptotics one has to be careful with choosing one of the Mittag-Leffler functions as an object suitable to replace the KWW function - to make the proper choice it is necessary to have information concerning the middle and long time behavior of the relaxation function. 

\section{Conclusions}\label{sec6}

It is known that the KWW function does not describe properly many relaxation phenomena, much better (and still friendly in use) is to use the CC model. But also this model breaks down in many physically interesting case - this was the reason of introducing the HN and JWS models which as phenomenological schemes fitting the data in the frequency domain. If transformed to the time domain both these models lead to the time decay laws given in terms of multiparameter Mittag-Leffler functions, long time unfamiliar to the physicists' community. Simultaneously to get information on the time decay of dielectric polarization is often much more needed for practical applications than data obtained the spectroscopy experiments although the latter are more precise and cover much larger range of the frequency involving 10 or even more orders of magnitude. Unfortunately, a fear of using unpopular and scary looking special functions of the Mittag-Leffler family effectively discourages a vast majority of physicists (first of all the experimentalists) and causes that they consider fitting the data by the stretched exponential function not as a routine coming from the long-time habit but as a method which is the only doable procedure. We consider this situation perplexing  and propose to give up this long-time habit. Having in mind that functions describing the relaxation phenomena are widely unknown we propose to begin with analysis of the characteristic exponents expressed in terms of easily calculable functions. Properties of these functions illustrate the problems which we face when compare various theoretical schemes used in the relaxation theory, in particular the problem of choosing the most suitable relaxation pattern. As a benchmark for comparing different relaxation patterns we took the stretched exponential and collated it, one by one, with well established models: Cole-Cole, Havriliak-Negami and Jurlewicz-Weron-Stanislavsky. These models are significantly different and overlap only in the asymptotic regime of large $s$, i.e., short times. Nevertheless we do not consider this result valueless - it gives a warning that to choose properly we have to look for additional information, e.g. the higher order short time asymptotics of relaxation functions or their long-time behavior. Tools to be used in order to push forward such investigations are collected and listed in the Sec. \ref{sec5}. Thus our main conclusion is that both theoretical studies as well as the time domain measurements of polarization decays should go beyond the stretched exponential fit and should use more extensively the results obtained by spectroscopy methods translated to the time domain. Here we would like to emphasize that Mittag-Leffler functions are well manageable with standard computer mathematics packages, like Mathematica, Matlab and Maple, and it is not a great problem to familiarize with them.

%%%%%%%%%%%%%%%%%%%%%%%%%%%%%%%%%%%%%%%%%%
%\vspace{6pt} 
%
%%%%%%%%%%%%%%%%%%%%%%%%%%%%%%%%%%%%%%%%%%%
%%% optional
%%\supplementary{The following are available online at \linksupplementary{s1}, Figure S1: title, Table S1: title, Video S1: title.}
%
%% Only for the journal Methods and Protocols:
%% If you wish to submit a video article, please do so with any other supplementary material.
%% \supplementary{The following are available at \linksupplementary{s1}, Figure S1: title, Table S1: title, Video S1: title. A supporting video article is available at doi: link.}
%
%%%%%%%%%%%%%%%%%%%%%%%%%%%%%%%%%%%%%%%%%%%
%\authorcontributions{Conceptualization, K.G., A.H., and K.A.P.; methodology, K.G., A.H., and K.A.P.; formal analysis, K.G., A.H., and K.A.P.; investigation, K.G., A.H., and K.A.P.; writing--original draft preparation, K.G., A.H., and K.A.P.; writing--review and editing, K.G., A.H., and K.A.P. All authors have read and agreed to the published version of the manuscript.}

%%%%%%%%%%%%%%%%%%%%%%%%%%%%%%%%%%%%%%%%%%
\section*{Acknowledgments}

K.G. and A.H. have been supported by the Polish National Center for Science (NCN) research grant OPUS12 no. UMO-2016/23/B/ST3/01714. K. G. acknowledges also support under the project Preludium Bis 2 no. UMO-2020/39/O/ST2/01563 awarded by the NCN and NAWA (Polish National Agency For Academic Exchange).

%%%%%%%%%%%%%%%%%%%%%%%%%%%%%%%%%%%%%%%%%%
%\acknowledgments{In this section you can acknowledge any support given which is not covered by the author contribution or funding sections. This may include administrative and technical support, or donations in kind (e.g., materials used for experiments).}

%%%%%%%%%%%%%%%%%%%%%%%%%%%%%%%%%%%%%%%%%%
%\conflictsofinterest{The authors declare no conflict of interest.} 
%
%%%%%%%%%%%%%%%%%%%%%%%%%%%%%%%%%%%%%%%%%%%
%%% Only for journal Encyclopedia
%%\entrylink{The Link to this entry published on the encyclopedia platform.}
%
%%%%%%%%%%%%%%%%%%%%%%%%%%%%%%%%%%%%%%%%%%%
%%% Optional
%\abbreviations{The following abbreviations are used in this manuscript:\\
%
%\noindent 
%\begin{tabular}{@{}ll}
%KWW & Kohlrausch--Williams--Watts relaxation \\
%CC & Cole--Cole relaxation\\
%CD & Cole-Davidson relaxation \\
%HN & Havriliak--Negami relaxation\\
%JWS & Jurlewicz--Weron--Stanislavsky relaxation\\
%CMF & completely monotone function\\
%SF & Stieltjes function\\
%CBF & completely Bernstein function
%\end{tabular}}
%
%%%%%%%%%%%%%%%%%%%%%%%%%%%%%%%%%%%%%%%%%%%
%%% Optional
%\appendixtitles{no} % Leave argument "no" if all appendix headings stay EMPTY (then no dot is printed after "Appendix A"). If the appendix sections contain a heading then change the argument to "yes".
\appendix

\section{The stretched exponential and Mittag-Leffler functions}\label{a0}

The L\'{e}vy stable distributions take distinguished place in our considerations \footnote{We shall exploit extensively the new class of L\'{e}vy distributions for rational $\alpha<1$ which exact and explicit forms was found in \cite{KAPenson10}.} because they enter the integral representations of basic functions used to describe non-Debye relaxations, namely the stretched exponential $\exp(- a p^{\alpha})$ \cite{HPollard46} and the one-parameter Mittag-Leffler function $E_{\alpha}(-a p^{\alpha})$ \cite{HPollard48, KWeron96}, both with $a > 0$ and $\alpha\in(0, 1]$. Introducing for $x, y > 0$ modified  functions $g_{\alpha}(x, y) = x^{-1/\alpha} g_{\alpha}(y\, x^{-1/\alpha})$ we rewrite the standard Pollard definitions \cite{HPollard46, HPollard48} as
\begin{multline}\label{30/07-3}
\E^{-a p^{\alpha}} = \int_{0}^{\infty} \E^{-u p} g_{\alpha}(a, u) \D u \quad \text{and} \\ E_{\alpha}(-a p^{\alpha}) = \int_{0}^{\infty}\E^{-u a} \frac{p}{\alpha u} g_{\alpha}(u, p) \D u. 
\end{multline}
Equations \eqref{30/07-3} have the form of the Laplace integrals but the variable $u$ enters them in different ways: either through $g_{\alpha}(a, u) = a^{-1/\alpha} g_{\alpha}(u a ^{-1/\alpha})$ or through $g_{\alpha}(u, p) = u^{-1/\alpha} g_{\alpha}(p u^{-1/\alpha})$. Worthy to note is also the different shape of differential relations held for the stretched exponential and the one-parameter Mittag-Leffler function 
\begin{multline}\label{24/08-1}
\frac{\D}{\D p} \E^{-a p^{\alpha}} = -a \alpha p^{\alpha - 1} \E^{-a p^{\alpha}} \quad \text{and} \\ {_{c}D_{p}^{\alpha}} E_{\alpha}(-a p^{\alpha}) = - a E_{\alpha}(-a p^{\alpha}).
\end{multline}
The first equality above is a differential relation which introduces the so-called Weibull distribution $\alpha p^{\alpha - 1} \exp(-a p^{\alpha})$ \cite{WWeibull51} while in the second equality we deal with an integro-differential relation of the eigenequation form in which the operator ${_{c}D^{\alpha}_{p}}$ denotes the Caputo fractional derivative, see Appendix \ref{a5}.  The one-parameter Mittag-Leffler function 
\begin{multline}\label{8/11-1}
E_{\alpha}(-x) = 1 - \frac{x}{\Gamma(1+\alpha)} + \frac{x^{2}}{\Gamma(1 + 2\alpha)} + \ldots \\= \sum_{r\geq 0} \frac{(-x)^{r}}{\Gamma(1 + \alpha r)}
\end{multline}
constitutes a generalization of exponential function as its series expansion differs from the usual exponential function series only by a parameter $\alpha$ present in the argument of the $\Gamma$ function settled in denominator. Moreover, the $\alpha$ parameter indicates that the eigenequation in  Eq. \eqref{24/08-1} is a generalization of the differential equation for the exponential function obtained when we put $\alpha=1$ in any of the equations Eq. \eqref{24/08-1}. 

The one-parameter Mittag-Leffler function provides us with the representative of a class of functions which generalize the exponential function and are widely met in studies of anomalous kinetic phenomena. Another frequently used function belonging to this class is the three-parameter Mittag-Leffler function $E_{\alpha, \beta}^{\gamma}(x)$ whose series form is 
\begin{multline*}
E_{\alpha, \beta}^{\gamma}(-x) = {1 - \frac{\gamma x}{\Gamma(\beta + \alpha)} + \frac{\gamma(\gamma + 1) x^{2}}{2\Gamma(\beta + 2\alpha)} - \ldots} \\= \frac{1}{\Gamma(\gamma)}\sum_{r\geq 0} \frac{\Gamma(\gamma + r) (-x)^{r}}{r!\Gamma(\beta + \alpha r)}, \quad \alpha, \beta, \gamma > 0 
\end{multline*}
while its integral representation reads
\begin{multline}\label{30/07-5}
E_{\alpha, \beta}^{\gamma}(-a p^{\alpha}) = \int_{0}^{\infty} \E^{-u a} \frac{p}{\alpha u} g_{\alpha, \beta}^{\gamma} (u, p)\D u, \\ g_{\alpha, \beta}^{\gamma}(u, p) = u^{-1/\alpha} g_{\alpha, \beta}^{\gamma}(p u^{-1/\alpha}),
\end{multline}
with a generalization of the one-sided L\'{e}vy stable distribution $g_{\alpha}(u, p)$, denoted by $g_{\alpha, \beta}^{\gamma}(u, p)$, being used \cite{KGorska21}. For rational $\alpha = l/k\in(0, 1)$ the function $g_{\alpha, \beta}^{\gamma}(u, p)$ can be expressed  through the Meijer $G$ function (see \cite[Corollary 3]{KGorska21} and Appendix \ref{a1})
\begin{equation}\label{30/07-6}
g_{\alpha, \beta}^{\gamma}(x) = \frac{1}{\Gamma(\gamma)} \frac{l^{1-\beta}}{k^{1-\gamma}} \frac{\sqrt{l k}}{(2\pi)^{(k-l)/2}} \frac{1}{x} G^{k, 0}_{l, k}\Big(\frac{l^{l}}{k^{k} x^{l}}\Big|{\Delta(l, \beta) \atop \Delta(k, \gamma)}\Big).
\end{equation}
For $\alpha = l/k$ and $\beta = \gamma = 1$ Eq. \eqref{30/07-6} reduces to the one-sided L\'{e}vy stable distribution $g_{l/k, 1}^{1}(x) = g_{l/k}(x)$ so we can say that the three-parameter Mittag-Leffler function turns into the one-parameter Mittag-Leffler function for $\alpha\in(0, 1)$ and $\beta = \gamma = 1$. For $\alpha, \beta\in(0, 1)$ and $\gamma = 1$ the three-parameter Mittag-Leffler function boils down to the two-parameter Mittag-Leffler function bearing the name of Wiman \cite{AWiman05}. One of the most useful properties of the three-parameter Mittag-Leffler function multiplied by a monomial $x^{\beta - 1}$,  in the literature called the Prabhakar function, 
is that its Laplace transform takes the form of simple rational function  
\begin{equation}\label{25/08-1}
\mathscr{L}[x^{\beta - 1}E_{\alpha, \beta}^{\gamma}(-a x^{\alpha}); s] = \frac{s^{\alpha\gamma-\beta}}{(a + s^{\alpha})^{\gamma}}.
\end{equation}

\section{The Fox $H$, Meijer $G$, and generalized hypergeometric functions}\label{a1}

The Fox $H$ function is defined by a Mallin-Barnes  contour integral, see  Ref. \cite{APPrudnikov-v3}:
\begin{multline}\label{31/08-1}
H_{p, q}^{m, n}\left(z\Big\vert { [a_{p}, A_{p}] \atop [b_{q}, B_{q}]}\right) \\= \frac{1}{2\pi\!\I} \int_{L_{H}} \frac{\prod_{j=1}^{m} \Gamma(b_{j} + B_{j}s)\, \prod_{j=1}^{n} \Gamma(1 - a_{j} - A_{j}s)}{\prod_{j=n+1}^{p} \Gamma(a_{j} + A_{j}s)\, \prod_{j=m+1}^{q} \Gamma(1 - b_{j} - B_{j}s)} z^{-s} \D s,
\end{multline}
where empty products are taken to be equal to one. In Eq. \eqref{31/08-1} the parameters are subject of conditions
\begin{align}\label{31/08-2}
\begin{split}
& z \neq 0, \quad 0 \leq m \leq q, \quad 0 \leq n \leq p; \\
& A_{j} > 0, \quad a_{j} \in\mathbb{C}, \quad j = 1, \ldots, p; \\ & B_{j} > 0, \quad b_{j} \in\mathbb{C}, \quad j = 1, \ldots, q; \\
& [a_{p}, A_{p}] = (a_{1}, A_{1}), \cdots (a_{p}, A_{p}); \\ & [b_{q}, B_{q}] = (b_{1}, B_{1}), \cdots (b_{q}, B_{q}).
\end{split}
\end{align}
For $A_{j} = B_{j} = 1$ the Fox $H$ functions reduces to the Meijer $G$ function:
\begin{multline}\label{31/08-3}
G_{p, q}^{m, n}\left(z\Big\vert { (a_{p}) \atop (b_{q})}\right)\\ = \frac{1}{2\pi\!\I} \int_{L_{G}} \frac{\prod_{j=1}^{m} \Gamma(b_{j} + s)\, \prod_{j=1}^{n} \Gamma(1 - a_{j} - s)}{\prod_{j=n+1}^{p} \Gamma(a_{j} + s)\, \prod_{j=m+1}^{q} \Gamma(1 - b_{j} - s)} z^{-s} \D s,
\end{multline}
where conditions listed in Eq. \eqref{31/08-2} become conditions for $[a_{p}, 1] = (a_{p}) = a_{1}, \cdots, a_{p}$ and $[b_{q}, 1] = (b_{q}) = b_{1}, \cdots, b_{q}$.
For a full description of the integration contour $L_{H}$ and $L_{G}$ and its properties as well as special cases for the $H$ and $G$ functions, see \cite[Secs. 8.2 and 8.3]{APPrudnikov-v3}. 

The generalized hypergeometric function ${_{p}F_{q}}$ is defined as follows
\begin{equation}\label{1/09-1}
{_{p}F_{q}}\left({a_{1}, \cdots, a_{p} \atop b_{1}, \cdots, b_{q}}; z\right) = \sum_{r\geq 0} \frac{z^{r}}{r!}\, \frac{\prod_{j=1}^{p} (a_{j})_{r}}{\prod_{j=1}^{q} (b_{j})_{r}},
\end{equation}
where $(a)_{r}$ is the Pochhammer symbol (rising factor) equals to $\Gamma(a + r)/\Gamma(a) = a (a+1) \cdots (a + r-1)$.

\section{The completely monotone and completely Bernstein functions}\label{a4}

The Bernstein functions (BFs) are non-negative functions on $\mathbb{R}_{+}$, differentiable there infinitely many times and satisfying for $s\in \mathbb{R}_{+}$ and $n\in\mathbb{N}_{0}$ the conditions $(-1)^{n}h^{(n+1)}(s)\ge 0$  everywhere in their domain.

The function $\widehat{h}(s)$, $s\in\mathbb{R}_{+}$, is a completely Bernstein function (CBF) if it is BF and $\widehat{h}(s)$ and $s/\widehat{h}(s)$ have the representation given by the Stieltjes transform \cite{RSchilling10}.\\
Alternative criterion says that $\widehat{h}(s)$ is CBF if $s/\widehat{h}(s)$ is CBF \cite{RSchilling10}.

{The completely monotone functions (CMF)} $\widehat{H}(s)$ are non-negative function of a non-negative argument whose all derivatives exist and alternate on $\mathbb{R}_{+}$, i.e.
\begin{equation*}
(-1)^{n} \widehat{F}^{(n)}(s) \geq 0, \quad n=0, 1, \ldots.
\end{equation*}

{The Bernstein theorem} uniquely and mutually connects CMF with the non-negative function defined on $\mathbb{R}_{+}$ by the Laplace transform:
\begin{equation*}
\widehat{H}(s) = \int_{0}^{\infty} \E^{-s t} F(t) \D t,
\end{equation*}
where $\widehat{H}(s)$ is CMF \cite{DVWidder46, HPollard44, ANKochubei11}. \\

\noindent
We can say that $f$ the Stieltjes function (SF) if, and only if, $1/f$ is a CBF \cite{RSchilling10}.  

\section{Relation between the infinitely divisible distribution and the Bernstein-class functions}\label{a3}

The relation between the CBF and infinitely divisible function is expressed by \cite[Lemma 9.2]{RSchilling10}. It say that the measure $g$ on $[0, \infty)$ is infinitely divisible iff $\mathcal{L}[g; \lambda] = \exp[-f(\lambda)]$ where $f$ is CBF.

\section{Fractional integrals and derivatives}\label{a5}

The fractional integral $(I^{\alpha}f)(x)$ for $\alpha\in(0, 1)$ equals to
\begin{equation*}
(I^{\alpha}f)(x) \okr \int_{0}^{x} f(\sigma) (x - \sigma)^{\alpha-1} \D\sigma/\Gamma(\alpha)
\end{equation*}
and fractional derivative in the Caputo sense $({_{c}D^{\alpha}}f)(x)$, also with $\alpha\in(0, 1)$, coincides with $(I^{1-\alpha} \frac{\D}{\D x} f)(x)$. {Explicitly for $\alpha\in(0, 1)$ it reads}
\begin{equation*}
(_{c}D^{\alpha} f)(x) \okr [\Gamma(1-\alpha)]^{-1} \int_{0}^{x} (x - y)^{-\alpha} f'(y)\, \D y
\end{equation*}
If $\alpha = 1$ reduces to the ordinary derivative: $\lim_{\alpha \to 1^{-}} (_{c}D^{\alpha} f)(x) = f'(x)$ \cite{IPodlubny99}. 

The pseudo--operators on the left hand side of Eqs. \eqref{9/08-1} and \eqref{9/08-2} belong to the class of Prabhakar--like integral operators which for the considered case are described in \cite[Appendix B]{RGarrappa16}
\begin{equation*}
{^{\rm C}}(D^{\alpha} + a)^{\beta} f(x)\okr \int_{0}^{x} (x - \sigma)^{-\alpha\beta} E_{\alpha, 1-\alpha\beta}^{-\beta}[-a(x-\sigma)^{\alpha}]\, f'(\sigma) \D\sigma 
\end{equation*}
and
\begin{equation*}
(D^{\alpha} + a)^{\beta} f(x)\okr \frac{\D}{\D\sigma} \int_{0}^{x} (x - \sigma)^{-\alpha\beta} E_{\alpha, 1-\alpha\beta}^{-\beta}[-a(x-\sigma)^{\alpha}] f(\sigma) \D\sigma. 
\end{equation*}
Eqs. \eqref{9/08-2} and \eqref{9/08-3} which should be completed with a suitable initial condition.

%%%%%%%%%%%%%%%%%%%%%%%%%%%%%%%%%%%%%%%%%%
%\section{References}

% Please provide either the correct journal abbreviation (e.g. according to the �??List of Title Word Abbreviations�?? http://www.issn.org/services/online-services/access-to-the-ltwa/) or the full name of the journal.
% Citations and References in Supplementary files are permitted provided that they also appear in the reference list here. 

%=====================================
% References, variant A: external bibliography
%=====================================
%\externalbibliography{yes}
%\bibliography{your_external_BibTeX_file}

\begin{thebibliography}{99}

\bibitem{FAlvarez91}
Alvarez, F., Alegr\'{i}a, A., Colmenero, J. Relationship between the time-domain Kohlrausch-Williams-Watts and frequency-domain Havriliak-Negami relaxation functions. {\em Phys. Rev. B} {\bf 1991} {\em 44}, 7306--7312.

\bibitem{FAlvarez93}
Alvarez, F.,  Alegr\'{i}a, A., Colmenero, J. Interconnection between frequency-domain Havriliak-Negami and time-domain Kohlrausch-Williams-Watts relaxation functions. {\em Phys. Rev. B} {\bf 1993} {\em 47}, 125--130.

\bibitem{AApelblat21}
Apelblat, A., Mainardi, F. Application of the Efros theorem to the function represented by the inverse Laplace transform of $s^{-\mu} \exp(-s^{\nu})$. {\em  Symmetry} {\bf 2021} {\em 13}, 354 (15pp).

\bibitem{ABaule03}
Baule, A., Friedrich, R. Joint probability distribution for a class on non-Markovian processes, {\em Phys. Rev.} {\bf 2003} {\em 71} 026101.

\bibitem{CJFBoettcher96}
B\"{o}ttcher, C. J. F., Bordewijk, P., Theory of electric polarization. vol. 2; Elsevier: Amsterdam, Holland, 1996.

\bibitem{AChechkin21} 
Chechkin, A.V., Sokolov, I.M. On relation between generalized diffusion and subordination schemes. {\em Phys. Rev. E} {\bf 2021} {\em103}, 032133 (10pp.).

\bibitem{AEfross35} 
Efross, A.M. The application of the operational calculus to the analysis. {\em Mat. Sb.} {\bf 1935} {\em 42}, 699--706, in Russian.

\bibitem{Erdelyi53}
Erd\'{e}lyi, A., Magnus, W., Oberhettinger, F., Tricomi, F.G., Higher Transcendental Function. vol.2; McGraw-Hill: New York,Toronto, London, 1953. 

\bibitem{HCFogedby94}
Fogedby, H.C. Langevin equations for continuous time L\'{e}vy flights. {\em Phys. Rev. E} {\bf 1994} {\em 50}, 1657--1660.

\bibitem{RGarrappa16} 
Garrappa, R., Mainardi, F., Maione, G. Models of dielectric relaxation based on completely monotone functions. {\em Frac. Calc. Appl. Anal.} {\bf 2016} {\em 19}, 1105--1160; corrected version available in {\tt arXiv: 1611.04028}

\bibitem{RTTGettens09}
{Gettens, R. T. T., Gilbert J. L. The electrochemical impedance of polarized 316L stainless steel: Structure-property-adsorption correlation. {\em J. Biomed. Mater. Res. A} {\bf 2008} {\em 90}, 121--132.}

\bibitem{RGorenflo14}
Gorenflo, R., Kilbas, A.A., Mainardi, F., Rogosin, S.V. Mittag-Leffler Functions, Related Topics and Applications: Theory and Applications. Springer: Berlin, Germany, 2014.

\bibitem{KGorska12a}
G\'{o}rska, K., Penson, K.A. L\'{e}vy stable distributions via associated integral transform. {\em J. Math. Phys.} {\bf 2012} {\em 53}, 053302 (10pp).

\bibitem{KGorska18} 
G\'{o}rska, K., Horzela, A., Bratek, {\L}., Penson, K.A., Dattoli, G. The Havriliak-Negami relaxation and its relatives: the response, relaxation and probability density functions. {\em J. Phys. A} {\bf 2018} {\em 51}, 135202 (15pp).

\bibitem{KGorska20} 
G\'{o}rska, K., Horzela, A. The Volterra type equation related to the  non-Debye relaxation. {\em Comm. Nonlinear Sci. Numer. Simulat.} {\bf 2020} {\em 85}, 105246 (14pp).

\bibitem{KGorska20a} 
G\'{o}rska, K., Horzela, A., Lattanzi, A., Pog\'{a}ny, T.K. On the complete monotonicity of the three parameter generalized Mittag-Leffler function $E_{\alpha, \beta}^{\gamma}(-x)$. {\em Appl. Anal. Discret. Math.} {\bf 2021} {\em 15}, 118--128. 

\bibitem{KGorska21} 
G\'{o}rska, K., Horzela, A. Non-Debye Relaxations: Two types of memories and their Stieltjes character, {\em Mathematics} {\bf 2021} {\em 9}, 477 (13pp). 

\bibitem{KGorska21a} G\'{o}rska, K., Horzela, A., Pog\'{a}ny, T.K. Non-Debye relaxations: smeared time evolution, memory effects, and the Laplace exponents, {\em Comm. Nonlinear Sci. Numer. Simulat.}  {\bf 2021} {\em 99}, 105837 (11pp).

\bibitem{KGorska21b}
G\'{o}rska, K. Integral decomposition for the solutions of the generalized Cattaneo equation. {\em Phys. Rev. E} {\bf 2021} {\em 104}, 024113 (11pp).

\bibitem{Ryzhik00}
Gradsteyn, I.S., Ryzhik, I.M., Tables of Integrals, Series and Products, 6th ed., Academic: San Diego, CA, 2000.

\bibitem{UGraf04}
Graf, U. Applied Laplace Transforms and $z$-Transforms for Sciences and Engineers. Birkh\"{a}user: Basel, Switzerland, 2004.

\bibitem{GGripenberg90} 
Grippenberg, G., Londen, S.O., Staffans, O.J. Volterra Integral and Functional Equations. Cambridge University Press: Cambridge, UK; 1990

\bibitem{MHaeri11}
{Haeri, M., Goldberg, S., Gilbert J. L. The voltage-dependent electrochemical impedance spectroscopy of CoCrMo medical alloy using time-domain techniques: Generalized Cauchy?Lorentz, and KWW?Randles functions describing non-ideal interfacial behaviour. {\em corros. Sci.} {\bf 2011} {\em 53}, 582--588.}

\bibitem{AHanyga20}
Hanyga, A. A comment on a controversial issue: a generalized fractional derivative cannot have a regular kernel, {\em Frac. Calc. Appl. Anal.} {\bf 2020} {\em 23}, 211.

\bibitem{HavriliakHavriliak96}
Havriliak Jr, S, Havriliak, S. J. Comparison of the Havriliak-Negami and stretched exponential functions. {\em Polymers} {\bf 1996} {\em 37}, 4107--4110

\bibitem{EHernandez17}
{Hern\'{a}ndez-Balaguera, E., Polo, J. L. A generalized procedure for the coulostatic method using a constant phase element. {\em Electro. Acta} {\bf 2017} {\em 233}, 167--172.}

\bibitem{EHernandez20}
{Hern\'{a}ndez-Balaguera, E., Polo, J. L. On the potential-step hold time when the transient-current response exhibits a Mittag-Leffler decay. {\em J. Electro. Chem.} {\bf 2020} {\em 856}, 113631 (6pp).}

\bibitem{EHernandez21a}
{Hern\'{a}ndez-Balaguera, E. Coulostatics in bielectrochemistry; A physical interpretation of the electrode-tissue processes from the theory of fractional calculus. {\em Chaos, Solitons and Fractals} {\bf 2021} {\em 145} {110768}(8 pp)}

\bibitem{EHernandez21b}
{Hern\'{a}ndez-Balaguera, Numerical approximations on the transient analysis of bioelectric phenomena at long time scales. {\em Chaos, Solitons and Fractals} {\bf 2021} {\em 145} {110787}(8 pp)}

\bibitem{EHernandez21c}
{Hern\'{a}ndez-Balaguera, del Pozo, G., Arredondo, B., Romero, B., Pereyra, C., Xie, H., Lira-Cat\'{u}, M. Unraveling the Key Relationship Between Perovskite Capacitive Memory, Long Timescale Cooperative Relaxation Phenomena, and Anomalous J-V Hysteresis. {\em Solar RRL} {\bf 2021} {\em 5}, 2000707 (8pp).}

\bibitem{RHilfer02} 
Hilfer, R. Analytical representations for relaxation functions of glasses. {\em J. Non-Cryst. Solids} {\bf 2002} {\em 305}, 122--126.

\bibitem{RHilfer02a} 
Hilfer, R. $H$-function representations for stretched exponential relaxation and non-Debye susceptibilities in glassy systems. {\em Phys. Rev. E} {\bf 2002} {\em 65}, 061510 (5pp).

\bibitem{RHilfer02b} 
Hilfer R. Fitting the excess wing in the dielectric $\alpha$-relaxation of propylene carbonate. {\em J Phys: Condens Matter} {\bf 2002} {\em 14}, 2297 --2301. 

\bibitem{RHilfer02c} 
Hilfer R. Experimental evidence for fractional time evolution in glass forming materials. {\em Chem Phys} {\bf 2002} {\em 284}, 399 --408.

\bibitem{Johnston06}
Johnston, D.C. Stretched exponential relaxation arising from continuous sum of exponential decays. {\em Phys. Rev. B} {\bf 2006} {\em 74} 184430 (7 pp). 

\bibitem{AKJonscher92}
Jonscher, A.K. The universal dielectric response and its physical significance. {\em IEEE Trans. Electr. Insul.} {\bf 1992} {\em 27}, 407--423.

\bibitem{ANKochubei11} 
Kochubei, A.N. General fractional calculus, evolution equations, and renewal processes. {\em Integr. Eq. Oper. Theory} {\bf 2011} {\em 71}, 583--600.

\bibitem{RKLuneburg66}
Luneburg, R.K. Mathematical Theory of Optics. University of California Press: Berkeley, Los Angeles, USA, 1966; Chap. II.

\bibitem{FMainardi14}
{Mainardi, F. On some properties of the Mittag-Leffler function $E_{\alpha}(-t^{\alpha})$, completely monotone for $t > 0$ with 0 < $\alpha < 1$. {\em Discrete. Contin. Dyn. Syst. Ser. B} {\bf 2014} {\em 19}, 2267--2278.}

\bibitem{AJMakowski09}
Makowski, A.J., G\'{o}rska, K. Quantization of the Maxwell fish-eye problem and the quantum-classical correspondence. {\em Phys. Rev. A} {\bf 2009} {\em 79}, 052116 (6pp.).

\bibitem{JCMaxwell52}
Maxwell, J.C. The Scientific Papers. Dover: New York, USA, 1952.

\bibitem{RMetzler02}
{Metzler, R., Klafter, J. From stretched exponential to inverse power-law: fractional dynamics, Cole-Cole relaxation processes, and beyond. {\em J. Non-Cryst. Solids} {\bf 2002} {\em 305}, 81--87.}

\bibitem{KAPenson10}
Penson, K.A., G\'{o}rska, K. Exact and explicit probability densities for one-sided L\'{e}vy stable distributions. {\em Phys. Rev. Lett.} {\bf 2010} {\em 105}, 210604 (4pp).

\bibitem{IPodlubny99} Podlubny, I. Fractional Differential Equations, Academic Press: San Diego, USA, 1999.

\bibitem{HPollard44}
Pollard, H. The Bernstein-Widder theorem on completely monotonic functions. {\em Duke Math. J.} {\bf 1944} {\em 11},  427--430.

\bibitem{HPollard46} Pollard, H. The representation of $\E^{-x^{\lambda}}$ as a Laplace integral. {\em Bull. Amer. Math. Soc.} {\bf 1946} {\em 52}, 908--910.

\bibitem{HPollard48}
Pollard, H. The completely monotonic character of the Mittag-Leffler function $E_{\alpha}(-x)$. {\em Bull. Amer. Math. Soc.} {\bf 1948} {\em 54} 1115--1116.

\bibitem{APPrudnikov-v3}
Prudnikov, A., Brychkov, Yu., Marichev, O. Integrals and Series. More special functions. Gordon and Breach: New York, USA, 1990; vol. 3.

\bibitem{APPrudnikov-v1}
Prudnikov, A., Brychkov, Yu., Marichev, O. Integrals and Series. Elementary Functions. Gordon and Breach: New York, USA, 1998; vol. 1.

\bibitem{Schilling} 
Schilling, R.L. An introduction to L\'evy and Feller processes. In: From L\'evy--type processes to parabolic SPDEs. Adv. Courses Math. Birkh\"{a}user -- Springer: Cham, Germany, 2016; pp. 1--126.

\bibitem{RSchilling10} 
Schilling, R.L, Song, R., Vondra\v cek,  Z. Bernstein Functions. De Gruyter: Berlin, Germany, 2010.

\bibitem[Sokolov(2002)]{IMSokolov02}
Sokolov, I.M. Solution of a class of non-Markovian Fokker-Planck equation. {\em Phys. Rev. E} {\bf 2002} {\em 66}, 041101 (5pp.).

\bibitem{AStanislavsky15}
Stanislavsky, A., Weron, K., Weron, A. Anomalous diffusion approach to non-exponential relaxation in complex physical systems. {\em Commun. Nonlinear. Sci. Numer. Simulat.} {\bf 2015}, {\em 24}, 117--126.

 \bibitem{AStanislavsky16} Stanislavsky, A., Weron, K. Atypical Case of the Dielectric Relaxation Responses and its Fractional Kinetic Equation. {\em Frac. Calc. Appl. Anal.} {\bf 2016} {\em 19}, 212--228.

\bibitem{AStanislavsky19}
Stanislavsky, A., Weron, K., Fractional-calculus tools applied to study the nonexponential relaxation in dielectrics in V. E. Tarasov (ed.) {\em Handbook of Fractional Calculus  with Applications. Volume 5. Applications in Physics, Part B}; De Gruyter: Berlin, 2019.

\bibitem{AStanislavsky21}
Stanislavsky, A., Weron, A. Duality in fractional systems. {\em Commun. Nonlinear. Sci. Numer. Simulat.} {\bf 2021}, {\em 101}, 105861.

\bibitem{WWeibull51}
Weibull, W. A Statistical Distribution Function of Wide Applicability, {\em J. App. Mech.-Trans. ASME} {\bf 1951} {\em 18}, 293--297.

\bibitem{KWeron96}
Weron, K., Kotulski, M. On the Cole-Cole relaxation function and related Mittag-Leffler distribution. {\em Physica A} {\bf 1996} {\em 232}, 180--188.

\bibitem{DVWidder46}
Widder, D.V. The Laplace Transform. Princeton University Press: London, UK, 1946.

\bibitem{AWiman05}
Wiman, A. \"{U}ber den Fundamentalsatz in der Theorie der Funktionen $E_{\alpha}(x)$. {\em Acta Math.} {\bf 1905} {\em 29}, 191--201.

\bibitem{LWlodarski52}
W{\l}odarski, \L. Sur une formule de Efross, {\em Studia Math.} {\bf 1952} {\em 13}, 183--187.

\end{thebibliography}

%=====================================
% References, variant B: internal bibliography
%=====================================

\end{document}